\documentclass[aps,prd,twocolumn,amsmath,superscriptaddress,amssymb,showpacs,floatfix,nofootinbib, superscriptaddress, longbibliography]{revtex4-1}

\usepackage{graphicx}
\usepackage{dcolumn}
\usepackage{xcolor}
\usepackage{bm}
\usepackage{natbib}
\usepackage{calligra}
\usepackage[T1]{fontenc}
\usepackage{egothic}
\usepackage[T1]{fontenc}
\newfont{\rsfsten}{rsfs10 scaled 1200}
\newfont{\rsfsseven}{rsfs10 scaled 1200}
\newfont{\rsfsfive}{rsfs10 scaled 1200}
\usepackage{epsfig}
\usepackage{units}
\usepackage{hyperref}
\usepackage[utf8]{inputenc}
\usepackage{cals, calc}

\let\nc=\nullcell                                                  
\let\sc=\spancontent

\newcommand{\col}{\dimexpr(19mm)+9.4pt\relax} 
\newcommand{\colW}{\dimexpr(23mm)+9.4pt\relax}
\newcommand{\colN}{\dimexpr(15mm)+9.4pt\relax}

\usepackage{lipsum}

\begin{document}

\title{ 
Exploring the Spectral Shape of Gravitational Waves Induced by Primordial Scalar Perturbations and Connection with the Primordial Black Hole Scenarios
}

\author{Ioannis Dalianis}
\email{dalianis@mail.ntua.gr}
\affiliation{Physics Division, National Technical University of Athens, 15780 Zografou campus, Athens, Greece}
\author{Konstantinos Kritos}
\email{ge16004@central.ntua.gr}
\affiliation{Physics Division, National Technical University of Athens, 15780 Zografou campus, Athens, Greece}


\begin{abstract}

There is a growing expectation that the gravitational wave detectors will start probing the stochastic gravitational wave backgrounds in the following years.
We explore the spectral shapes of gravitational waves induced to second order by scalar perturbations and presumably have been produced in the early universe.
We calculate the gravitational wave spectra
generated during radiation and kination eras
together with 
the associated primordial black hole counterpart.
We employ power spectra for the primordial curvature perturbation generated by $\alpha$-attractors and nonminimal derivative coupling inflation models as well as Gaussian and delta-type shapes.
We demonstrate the ability of the tensor modes to constrain the spectrum of the primordial curvature perturbations and discriminate among inflationary models.
Gravitational wave production during kination and radiation eras can also be distinguished by their spectral shapes and amplitudes.

\end{abstract}

\maketitle

\section{Introduction}

 A network of operating and designed gravitational wave detectors raise expectations that, sooner or later, the relic gravitational radiation background will be detected. Gravitational waves (GWs) are thought to be generated by several processes in the early universe \cite{Kuroyanagi:2018csn, Christensen:2018iqi, Caprini:2018mtu}. Among them there is a model independent source, the primordial density scalar perturbations, that unambiguously source tensor modes at second order perturbation theory \cite{Matarrese:1992rp, Matarrese:1997ay, Noh:2004bc, Carbone:2004iv, Nakamura:2004rm}, the so-called induced gravitational waves (IGW).
 
Tensor modes are induced to second order by scalar perturbations despite their decoupling at first order.
The cosmological implications of the scalar induced GWs to second order were first discussed in the pioneering works \cite{Mollerach:2003nq}, where the effects of IGWs on cosmic microwave background (CMB) polarization was computed, and \cite{Ananda:2006af}, where the spectrum of the IGWs on small scales was studied.
Further notable results followed that extended our understanding and formulation of IGW physics \cite{ Assadullahi:2009nf, Baumann:2007zm, Saito:2008jc}
and gave us insights of how to probe
the primordial power spectrum, ${\cal P_R}(k)$, at small scales.

Primordial scalar density perturbations are directly measured only 
at the very large scales of the CMB with wave number
$k_\text{cmb}\sim 10^{-2}$ Mpc$^{-1}$ and with amplitude   $\delta\rho/\rho \sim 10^{-5}$.
At smaller scales $k\gg k_\text{cmb}$ there are weak constraints 
coming mainly from bounds on the abundance of primordial black holes (PBHs) \cite{Dalianis:2018ymb, Kalaja:2019uju, Carr:2020gox}. Additional constraints on the scalar density amplitude can come from gravitational wave experiments. Already, at the frequency range $f\sim 10^{-9}$ Hz and $f\sim 10^2$ Hz the PTA and LIGO/VIRGO experiments, respectively, have provided upper bounds.

The weak bounds on the primordial ${\cal P_R}(k)$ has prompted the building of several scenarios that exhibit an enhanced power at small scales. 
A subclass of these scenarios predicts a significant PBH abundance, see Ref. \cite{Sasaki:2018dmp} for a review.  The new observational window to the very early universe through the IGWs  can test these scenarios \cite{
Bugaev:2009kq, Saito_2010, Bugaev:2009zh, Bugaev:2010bb, Nakama:2016gzw, Orlofsky:2016vbd, Garcia-Bellido:2017aan, Cai:2018dig, Inomata:2018epa, Espinosa:2018eve, Bartolo:2018rku, Byrnes:2018txb, Bartolo:2018rku, Cai:2019jah,Wang:2019kaf, Gong:2019mui, Mahbub:2019uhl,  Ozsoy:2019lyy, Bartolo:2019zvb, Braglia:2020eai, Ballesteros:2020qam}.
The interpretation of a GW signal is nevertheless  a nontrivial task and one needs to take into account several parameters, such as the reheating temperature and the equation of state of the early universe, as well as whether the primordial fluctuations are Gaussian or a non-negligible non-Gaussianity exists \cite{Nakama:2016gzw, Cai:2018dig, Bartolo:2018rku, Ota:2020vfn}.
From a different point of view, detection of the relic GWs stochastic background is a direct probe of the very early  cosmic history, which is unknown for $t\lesssim 1$s \cite{Allahverdi:2020bys}.
In this context, the recent works \cite{Inomata:2020tkl, Inomata:2020yqv, Domenech:2019quo, Domenech:2020kqm} have studied the impact of different early cosmological evolution on the IGWs.

In this work we 
explore the shape of the IGWs produced in the early universe assuming different scenarios and models. Prior to BBN there are no direct tests of the thermodynamical state of the universe and we  choose to presume either the standard radiation equation of state or that of the stiff fluid, as benchmark cases. We make a step forward compared to previous works taking into account the transition details from kination to radiation. 
We focus on a ${\cal P_R}(k)$ shape with a peak.  ${\cal P_R}(k)$ with a peak is a rather interesting scenario because it predicts IGWs and might also generate PBHs that can be part of the dark matter in the galaxies. 
We target at cosmological parameter values that can be tested by the existing and designed GW experiments, such as LIGO and Laser Interferometer Space Antenna (LISA).
We utilize Dirac $\delta$ and Gaussian-type ${\cal P_R}(k)$ shapes and give a simple analytic expression for the energy density parameter of the IGWs, $\Omega_\text{IGW}(t_0)$, produced during radiation and kination regime. 

Furthermore, we derive the scalar-induced GW spectral shapes for particular  models of inflation: two $\alpha$-attractors inflation models \cite{Kallosh:2013yoa} and a generalized Galileon-type \cite{Horndeski:1974wa, Kobayashi:2019hrl, Deffayet:2011gz} model of inflation.
 We focus on particular three models \cite{Dalianis:2018frf, Dalianis:2019asr, Dalianis:2019vit} because  each model yields a ${\cal P_R}(k)$ with a different shape form: different width and scaling both in the IR and UV part of the scalar spectrum.  We  compare  features of the ${\cal P_R}(k)$ spectral shape with the corresponding $\Omega_\text{IGW}(f)$, examining the amplitudes and the scaling.  This approach makes possible a demonstrative exploration of the IGW spectra with different shapes. 
 A similar study in this direction has been performed in Refs. \cite{Cai:2019cdl, Pi:2020otn} where, following analytic steps, general features of the IGW spectrum have been also described.
 Our analysis provides observational tests for this sort of viable inflation models as well.

In addition to the IGW spectra we examine the abundance and distribution of the PBHs associated with the ${\cal P_R}(k)$. 
We examine scenarios that predict a sizable amount of both PBHs and IGWs,  and scenarios that predict significant IGWs with the PBH counterpart being either negligibly small or promptly evaporating.
We present analytic and numerical results and we manifest the connection between the PBH mass and the GW counterpart making use of figures and tables.

The paper is organized as follows. In Sec. \ref{SecMethod}, we review the method that is usually followed in the literature to calculate the spectra of scalar-induced GWs. In Sec. \ref{secEras} we focus on particular inflationary models that predict  ${\cal P_R}(k)$ with a peak as well as general Gaussian ${\cal P_R}(k)$ types, and we specialize into radiation and kination domination early universe scenarios.
In Sec. \ref{SecPBH} we turn into the PBH counterpart and calculate mass distributions and abundances. Our results are presented in Sec. \ref{SecResults}, and finally, in Sec. \ref{secConclusions} we draw our conclusions.

\section{Methodology for calculating Induced Gravitational Waves} \label{SecMethod}

In this section we review the formulas required to calculate the energy density of the induced gravitational waves (IGWs) by first order scalar perturbations and adjust the notation in our context. We follow Ref.~\cite{Kohri:2018awv} 
and for a further study in a relevant context we refer the reader to \cite{Kohri:2018awv, Domenech:2019quo, Pi:2020otn}.

We consider a flat Friedmann-Robertson-Walker model with first order scalar perturbations $\Phi$, $\Psi$ and second order (induced) tensor perturbations $h_{ij}$, and ignore vector perturbations and first order tensor perturbations. We describe our perturbations in the Newtonian gauge.
 The line element of the perturbed metric is written as,
\begin{multline}
    ds^2=a^2(\eta)\ \big[ -(1+2\Phi)\ d\eta^2
    \\
    +\big((1-2\Psi)\ \delta_{ij}+{1\over2} h_{ij}\big)\ dx^i\ dx^j\big],
\end{multline}
where $a(\eta)$ is the scale factor and $\delta_{ij}$ is the Kronecker tensor. For future reference we also denote the conformal Hubble function by $\mathcal{H}(\eta)\equiv{a{'}(\eta)\over a(\eta)}$. Here, and later, prime ($'$) denotes conformal time differentiation.
We will also ignore anisotropies\footnote{For the effects of the anisotropic stress due to  ${\cal O}(10)\%$  difference between $\Phi$ and $\Psi$  see Ref.~\cite{Baumann:2007zm}.} and set $\Phi=\Psi$. In the Newtonian gauge the scalar perturbations play the role of the gravitational potential and the tensor modes describe gravitational waves. 

We define $x\equiv k\eta$ and express the evolution of the scalar perturbations in terms of the scalar transfer function, $\Phi$, defined by $\Phi_{\textbf{k}}(\eta)\equiv \Phi(x)\ \phi_{\textbf{k}}$ in momentum space, where $\phi_{\textbf{k}}$ the Fourier mode of the primordial scalar perturbations. Considering the field equations of general relativity for a single perfect cosmological fluid with the standard density-pressure equation of state (EOS) relation $p=w \rho$, the evolution equations for the gravitational potential is obtained by 
\begin{eqnarray} 
\label{STFevol}
    \Phi{''}(x)+3(1+w)\ \mathcal{H}(\eta)  \Phi{'}(x)+w\ k^2\ \Phi(x)=0\,.
\end{eqnarray}
\\
We consider adiabatic perturbations (zero entropy perturbations) and initial conditions $\Phi(0)=1$ and $\Phi{'}(0)=0$ for the scalar transfer function. The general solution\footnote{ Here, we neglect the contribution from the Bessel function of the second kind, $Y_{\alpha}(\sqrt{w} x)$, to avoid the singularity at $\eta=0$.} of Eq.~(\ref{STFevol}) is given in terms of the Bessel function of the first kind, $J$, and the gamma function, $\Gamma$, \cite{Baumann:2007zm},
\begin{eqnarray} \label{EqPhiw}
    \Phi(x)={2^{\alpha}\ \Gamma(\alpha+1)\over \left(\sqrt{w} x\right)^{\alpha}}\,J_{\alpha}(\sqrt{w} x),
\end{eqnarray}\\
where $\alpha={5+3 w \over 2(1+3 w)}$.

The evolution of the induced tensor modes is given by the equation,
\begin{equation}
    h_{\textbf{k}}{''}(\eta)+2\mathcal{H}(\eta)h_{\textbf{k}}{'}(\eta)+k^2h_{\textbf{k}}(\eta)=\mathcal{S}_{\textbf{k}}(\eta),
\end{equation}
where the source function ${\cal S}_{\textbf{k}}$ at the left-hand side plays a critical role.  
It  is 
a convolution of scalar perturbations at different wave numbers
given by
\begin{eqnarray}
    \mathcal{S}_{\mathbf{k}}(\eta)=4\int\ {d^3q\over(2\pi)^{3/2}}\ e^{ij}(\textbf{k})\ q_i\ q_j\ \phi_{\textbf{k}}\ \phi_{\textbf{k}-\textbf{q}}
    \nonumber
    \\
    \times\ f(|\textbf{q}|/k,|\textbf{k}-\textbf{q}|/k,\eta,k).
\end{eqnarray}
Here, $e^{ij}$ is the polarization tensor, and $f$ is an auxiliary function defined by           
\begin{widetext}
\begin{equation}
    f(u,v,\eta,k)= 2\ \Phi(u x) \Phi(v x)
    +{4\over3(1+w)}\left[ \Phi(u x)+{u k\over\mathcal{H}(\eta)}\ \Phi{'}(u x)\right]
    \left[\Phi(v x)+{v k\over\mathcal{H}(\eta)}\ \Phi{'}(v x)\right].
\end{equation}
\end{widetext}

Gaussian fluctuations are best described  with the power spectral density. 
For scalar and tensor perturbations the power spectral densities, called $\mathcal{P}_{\Phi}$ and $\mathcal{P}_{h}$, respectively, are defined in terms  of two point correlation functions as
\begin{equation}
    \langle\phi_{\textbf{q}}\ \phi_{\textbf{k}}\rangle\equiv{2 \pi\over k^3}\ \mathcal{P}_{\Phi}(k)\ \delta^{(3)}(\textbf{q}+\textbf{k}),
\end{equation}
and 
\begin{equation}
    \langle h^{\oplus,\otimes}_{\textbf{q}}(\eta)\ h^{\oplus,\otimes}_{\textbf{k}}(\eta)\rangle\equiv{2 \pi\over k^3}\ \mathcal{P}_{h}(\eta,k)\ \delta^{(3)}(\textbf{q}+\textbf{k}),
\end{equation}
where $\delta^{(3)}$ is the three-dimensional delta distribution and $\oplus,\otimes$ denote the two polarization modes.
The power spectral density of the comoving curvature perturbation, $\mathcal{P}_{\mathcal{R}}$, is related to  $\mathcal{P}_{\Phi}$ via
\begin{equation}
    \mathcal{P}_{\mathcal{R}}(k)=\left({5+3 w\over3+3 w}\right)^2\ \mathcal{P}_{\Phi}(k).
\end{equation}

The density parameter of IGWs is defined to be the ratio of the energy density in IGWs over the total energy density, taking into account both $\oplus$ and $\otimes$ GW polarizations. 
A useful definition for the energy density of IGWs per unit logarithmic frequency interval  is given  in terms of the tensor power spectrum as
\begin{equation} \label{OmegaIGWc}
    \Omega_{\textrm{IGW}}(\eta,k)\equiv{1\over24}\ \left({k\over\mathcal{H}(\eta)}\right)^2\ \overline{\mathcal{P}_{h}(\eta,k)},
\end{equation}
where the overline denotes the oscillation average. The power spectral density of the tensor perturbations is expressed as a double integral involving the power spectrum of the curvature perturbations,
\begin{equation}
\label{tensorPSD}
    \overline{\mathcal{P}_{h}(\eta,k)}=\int_{0}^{\infty} dv\int_{|1-v|}^{1+v} du\ \mathcal{T}(u,v,\eta,k)\ \mathcal{P}_{\mathcal{R}}(u
k)\ \mathcal{P}_{\mathcal{R}}(v k).
\end{equation}
The $\mathcal{T}$ is the tensor transfer function given by
\begin{multline}
    \mathcal{T}(u,v,\eta,k)=4\ \left(4 v^2-(1+v^2-u^2)^2\over4 u v\right)^2
    \\
    \times \left({3+3w\over5+3w}\right)^2\ \overline{I^2(u,v,\eta,k)},
\end{multline}
where we have defined the kernel function by
\begin{equation}
    I(u,v,\eta,k)=\int_{0}^xdy\ {a(y/k)\over a(\eta)}\ kG_{k}(x,y)\ f(u,v,y/k,k),
\end{equation}
and have used the Green's function $kG_{k}$ given in terms of Bessel functions, as $kG_{k}(x,y)={\pi\over2} \sqrt{x\ y}\cdot \left[Y_{\nu}(x)\ J_{\nu}(y)-Y_{\nu}(y)\ J_{\nu}(x)\right]$, with $\nu\equiv{3(1-w)\over2(1+3 w)}$, \cite{Tomikawa:2019tvi}.
The oscillation average is given by
\begin{equation}
    \overline{I^2(u,v,\eta,k)}\equiv{1\over2\pi}\ \int_x^{x+2\pi} dy \ I^2(u,v,\eta,y/\eta).
\end{equation}

The IGWs are sourced gravitational waves.
At the time $t_\text{c}$ the production of IGWs ceases and afterward the GWs propagate freely. 
In a radiation dominated background the energy density parameter of the GWs also remains constant.
The energy density parameter of the IGW today, $t_0$, is therefore given by Eq.~(\ref{OmegaIGWc}) at the moment $t_\text{c}$ times the current radiation density parameter, $\Omega_{\gamma,0}h^2=4.2\times10^{-5}$, modulo changes in the number of the relativistic degrees of freedom $g_*$ in the radiation fluid,
\begin{multline}
\label{IGWtoday}    
    \Omega_{\textrm{IGW}}(t_0,f)h^2=0.39\times\left({g_{\ast}\over106.75}\right)^{-{1\over3}}\ \Omega_{\gamma,0}h^2\\\times\Omega_{\textrm{IGW}}(t_\text{c},f).
\end{multline}

For the total IGW density parameter, we integrate Eq. (\ref{IGWtoday}) over the logarithmic interval of frequency,
\begin{equation}
    \Omega_{\textrm{IGW}}(t_0)=\int_{f_\text{min}}^{f_\text{max}} d\ln f\ \Omega_{\textrm{IGW}}(t_0,f).
\end{equation}
Since our spectra will be peaked at some frequency $f_{\text{IGW}}^\text{p}$ we may take $f_\text{min}=f_{\text{IGW}}^\text{p}/100$ and $f_\text{max}=10f_{\text{IGW}}^\text{p}$. Here $\Omega_\text{IGW}(t_0)$ should not be confused with $\Omega_\text{IGW}(t_0,f)$ where the later function is frequency dependent and denotes the density parameter of IGWs per logarithmic frequency interval.

In the following we will examine particular ${\cal P_R}(k)$ types and we will focus on two benchmark early universe cosmological scenarios: the radiation and kination dominated eras.

\section{Study of the IGW$\text{s}$ produced by explicit models and in different cosmological eras} \label{secEras}

It is well known that the equation of state of the Universe has not been directly probed for times prior to BBN $t\sim 1$ s.
IGWs  have amplitude and spectral shape that depend on the details of cosmological era at the time of their production. Indeed, the $\Phi$ that sources the GWs evolves in a different way for different equation of state $w$ and, additionally, the energy density of the produced GWs scales differently with respect to the background. 

In order to study the IGW spectrum and demonstrate its properties and behaviour we assume that the curvature power spectrum ${\cal P_R}(k)$ features a peak at the wave number $k_\text{p}$. If the amplitude of the ${\cal P_R}(k)$ peak is significant it produces a strong  IGW signal, that is possibly detectable, and additionally, enhances exponentially the production probability for PBHs. Therefore, explicit models can be tested and upper bounds on the IGWs can be derived.

We will study ${\cal P_R}(k)$ peaks generated by complete and explicit inflationary models, namely the $\alpha$-attractors  and Horndeski  general nonminimal derivative coupling (GNMDC)  inflation introduced in Refs.~\cite{Dalianis:2018frf} and~\cite{Dalianis:2019vit} respectively. The  qualitative element of the first model is that features a near-inflection point  \cite{Garcia-Bellido:2017mdw, Germani:2017bcs, Motohashi:2017kbs, Ballesteros:2017fsr} and the second model features a so-called high friction regime \cite{Germani:2010gm}.
From the structural side the two inflationary models  \cite{Dalianis:2018frf, Dalianis:2019vit} have a noncanonical kinetic term,
\begin{align} 
\label{Lattr}
&\frac{{\cal L}^{\alpha-\text{attractors}}}{\sqrt{-g}}=\frac{R}{2}-\frac{(\partial_\mu \tilde{\varphi})^2}{2(1-\frac{\tilde{\varphi}^2}{6\alpha})^2}+V(\tilde{\varphi}) \\
& \frac{{\cal L}^\text{GNMDC}}{\sqrt{-g}}
=\frac{R}{2} -\hat{f}(\varphi) G^{\mu\nu}\partial_\mu \varphi \partial_\nu \varphi  +V(\varphi) \label{Lhorn}
\end{align}
We note that we examine the second order induced tensor perturbations, not the first order tensor modes produced by the inflationary stage itself; for the latter, see e.g., \cite{Kuroyanagi:2014qza,Bernal:2019lpc} for relevant studies.
What is of interest for our analysis is that, respectively, the form of the $\alpha$-attractor potential and the form of the GNMDC coupling can generate an enhanced ${\cal P_R}(k)$ spectrum at  $k_\text{p}$ and with a variety of shape forms.
The $\alpha$-attractors potentials that we utilize generate ${\cal P_R}(k)$ slopes that increase like $k^3$ and $k^4$ and decrease like $k^{-1}$ or $k^{-4}$. On the other hand the   ${\cal P_R}(k)$ slopes of the Galileon GNMDC model increase like $k^2$ and $k^4$ and decrease like $k^{-4}$.
For a comprehensive description and details about the inflationary Lagrangians~(\ref{Lattr}) and~(\ref{Lhorn}), as well as the generated ${\cal P}_R(k)$ shapes  we refer the reader to the original works \cite{Dalianis:2018frf, Dalianis:2019vit}. A brief description can be found in Appendix \ref{AInf}.

Moreover, irrespective of the inflationary theory, we assume general shapes for the ${\cal P_R}(k)$ spectra. We 
utilize Gaussian and Dirac $\delta$-type distributions,
\begin{align} \label{GaussD}
  &  {\cal P_R}^\text{G}(z)= A_{\cal R} \, e^{-(z/\epsilon)^2} = \frac{A_0}{\epsilon \sqrt{\pi}}e^{-(z/\epsilon)^2}
  \\
  & {\cal P_R}^\text{D}(z)= A_0 \, \delta(z) \label{deltaD}
\end{align}
where $z\equiv\ln(k/k_\text{p})$ and $A_{\cal R}={\cal P_R}(k_p)$. In the limit $\epsilon\rightarrow 0$ the Gaussian distribution approaches the $\delta$-distribution. 
The advantage of these distributions is that analytic or semianalytic results can be obtained and, in addition, realistic models can be parametrized. Gaussian spectra have been also studied recently in \cite{Pi:2020otn}.

We are interested in three characteristics for the power spectra ${\cal P_R}(k)$: i) the amplitude $A_{\cal R}$; ii) the wave number of the peak, $k_\text{p}$, where ${\cal P_R}(k_\text{p})=A_{\cal R}$; and iii) the width $\epsilon$.
 We choose a large amplitude $A_{\cal R}$ so that the IGW signal is significant. 
 We slightly modulate the $A_{\cal R}$ in order to maximize or minimize to a negligible amount the PBH abundance. The wave number $k_\text{p}$ is chosen so that either  the PBHs constitute a significant fraction of the dark matter in the galaxies, or the frequency of the IGWs lays in the sensitivity range of the gravitational detectors.

Let us now examine two benchmark cosmological scenarios separately: the radiation and the kination.

\subsection{The radiation domination scenario}

Let us assume that the 
${\cal P_R}(k)$ peak
enters the horizon during the radiation era. 
This is the standard early universe cosmological scenario. The IGWs produced during radiation era have been thoroughly studied, thus, we will not repeat known results and technical details. We remind the reader that the potential $\Phi(x)$ oscillates with a $x^{-2}$ decaying amplitude in subhorizon scales. 
The formalism of the Sec.~\ref{SecMethod} is applied for $w=1/3$.
We quote the analytic expression 
that we use to obtain the IGWs results for the models (\ref{Lattr}), (\ref{Lhorn}) and (\ref{GaussD}).
Relying on analytical methods, we obtain the tensor transfer function in late times as was done in \cite{Kohri:2018awv}; the resulting expression is

\begin{widetext}
\begin{eqnarray}
\label{radiation_TTF}
    \lim_{x\to\infty} x\ \mathcal{T}_{\textrm{RD}}(u,v,x)=\ 2\left({4v^2 - (1+v^2-u^2)^2 \over 4uv}\right)^2
    \left({3(u^2+v^2-3) \over 4u^3v^3}\right)^2
    \left[\left(-4uv + (u^2+v^2-3)\ln\left|{3-(u+v)^2 \over 3-(u-v)^2}\right|\right)^2  \right.  
    \nonumber
    \\
    \left.+ \pi^2 (u^2+v^2-3)^2 \Theta\left(u+v\sqrt{3}\right) \right] \ \ \ \ \ 
\end{eqnarray}
\end{widetext}
where $\Theta$ is the unit step function.

We then perform a numerical integration to calculate the ${\cal P}_h(\eta, k)$ as Eq.~(\ref{tensorPSD}) dictates.
In order for this to be done, an input for the power spectrum  shape
${\cal P_R}(k)$ 
is required. 
We assume two Gaussian power spectra with medium and narrow widths, $\epsilon=1$ and 
$\epsilon=0.1$, respectively, as well as power spectra produced by $\alpha$-attractor and Horndeski-type inflationary theories; see Fig.~\ref{Figps}.
The ${\cal P_R}(k)$ produced by these inflationary theories is found after solving the Mukhanov-Sasaki equation numerically.

\subsection{The kination domination scenario}
\label{subsec:kination}

In this subsection, let us assume the cosmological scenario that the very early universe has been dominated by a phase whose equation of state is stiffer than radiation. 
Kination is a regime where the kinetic energy of a scalar field is dominant against its potential energy. In the limiting case $w=1$, where the sound velocity is equal to the speed of light, the energy density of the stiff fluid scales as $\rho\sim a^{-6}$.
Such a scenario is natural in theories with runaway potentials, e.g. theories with moduli fields. 

We examine the kination scenario because it has  striking implications both for the PBH formation and the IGW produced. 
Our numerical results complement previous studies \cite{Domenech:2019quo} and we show that an early universe kination domination (eKD) era
shapes in a different way the spectrum of the tensor perturbations. This fact renders the eKD era testable.

We comment that the attribute "early" to kination might sound redundant since there is no kination era observed in the universe. Nevertheless we use it to emphasize that the kination era assumed precedes the standard radiation era.

 The redshift of the stiff fluid energy density is the fastest and any ambient radiation will sooner or latter dominate the early universe. Hence, a kination era ends naturally if there is an extra fluid with softer equation of state. Besides this gradual transition, an eKD era can end suddenly. 
 The later is simpler to examine because the scalar transfer function can be computed explicitly due to the discrete $w$ values and single fluid analysis.\footnote{
 A mechanism that can implement a sudden transition can be easily conceived: an extra field direction that ends the kination stage via a waterfall transition to a global minimum where the field decays and reheats the universe.}

Let us consider an early kination domination  scenario that transits suddenly into the RD phase.
The transition takes place at the reheating temperature $T_\text{rh}$. 
 The transfer function in the kination domination
cannot be written in closed analytical expressions.
One can only rely on approximate methods, since the Bessel functions of order one involved can only be written as infinite series and not further simplified. Hence, we treat the scenario eKD to RD numerically. 
Further, in order to simplify our calculations we utilize a monochromatic power spectrum of scalar perturbations modeled by a delta distribution peaked at $k_\text{p}$, given by Eq.~(\ref{deltaD}).
Such a choice models a narrow and sharply peaked $\mathcal{P}_{\mathcal{R}}$. In the conclusions, we will discuss a correspondence between $\delta$-type and Gaussian distributions, and  anticipate features for the IGWs produced by a wider ${\cal P_R}(k)$.

\begin{figure}
    \centering
    \includegraphics[width=8.2cm,height=5cm]{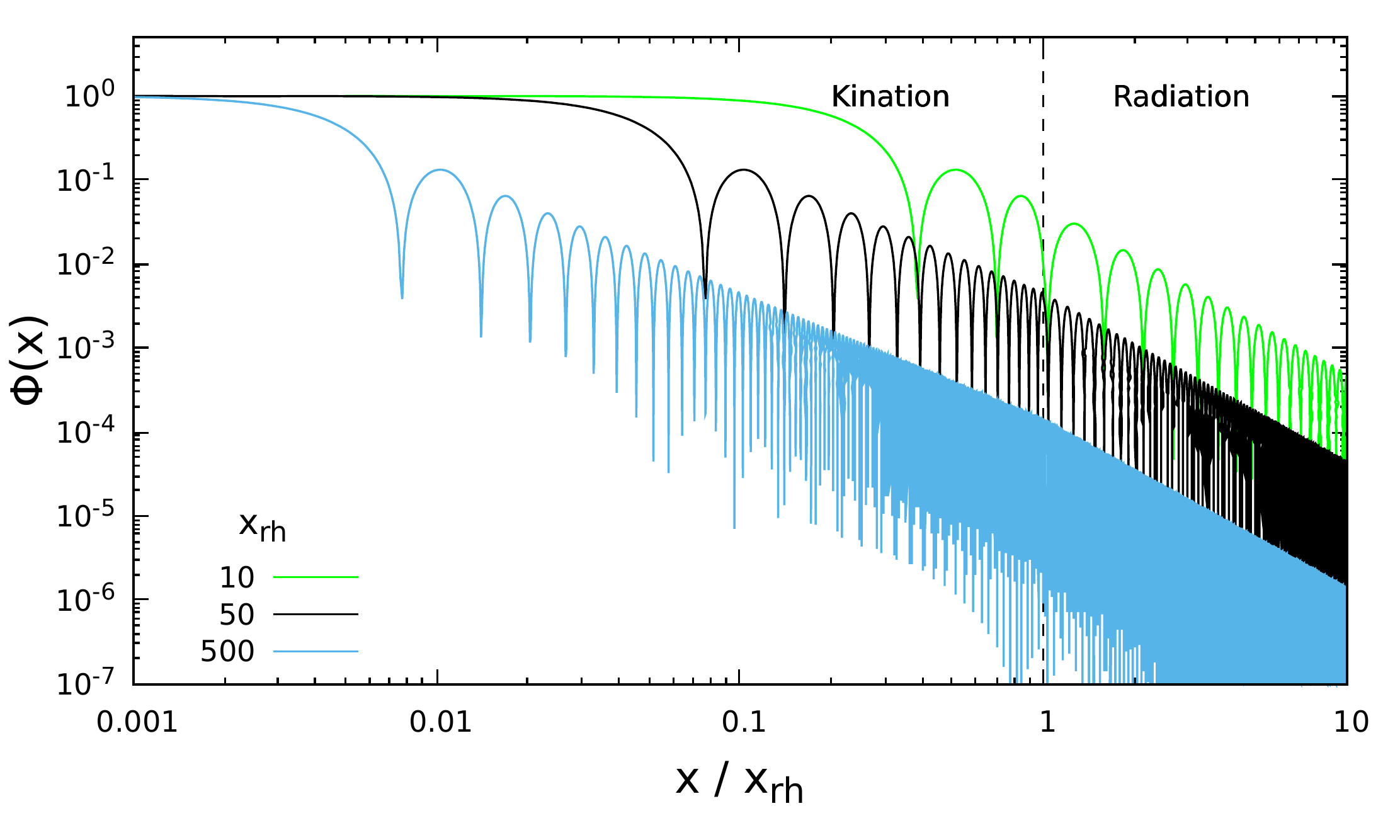}
    \caption{The transfer function in the case of the eKD$\rightarrow$RD sudden transition in the Newtonian gauge. We plot the absolute value of the gravitational potential, Eq.~(\ref{pot}),  for $x_\text{rh}=10,\ 50,\ 100$. The horizontal axis is normalized so that the transition occurs at 1.}
    \label{potentialPlot}
\end{figure}

In a sudden eKD$\rightarrow$RD scenario, the scale factor is a piecewise defined function, that is continuous at the point of transition $\eta=\eta_\text{rh}$,
\begin{equation}
    a(\eta)=\begin{cases}
    \displaystyle
    \sqrt{\eta\over\eta_\text{rh}}, & \eta<\eta_\text{rh}  \\
    \displaystyle
    {\eta+\eta_\text{rh}\over2\eta_\text{rh}}, & \eta\ge\eta_\text{rh}
    \end{cases}
\end{equation}
A similar expression can be obtained for the conformal Hubble parameter, ${\cal H}$. 
The scalar transfer function, Eq. (\ref{STFevol}), is given by
\begin{widetext}
\begin{equation}
\label{pot}
    \Phi(x)=\begin{cases}
    \displaystyle
    {2\over x}\ J_{1}(x), & x<x_\text{rh}\\
    \displaystyle
    {3\over x^2}\left[ C_1 \left({\sin(x/\sqrt{3})\over x/\sqrt{3}} - \cos(x/\sqrt{3})\right) + C_2 \left({\cos(x/\sqrt{3})\over x/\sqrt{3}} + \sin(x/\sqrt{3})\right) \right], & x\ge x_\text{rh}
    \end{cases}
\end{equation}
\end{widetext}
 After the transition, into the RD era, we write  the solution to  Eq.~(\ref{STFevol}) as a linear combination of $(x/\sqrt{3})^{3/2}\cdot J_{3/2}(x/\sqrt{3})$ and $(x/\sqrt{3})^{3/2}\cdot Y_{3/2}(x/\sqrt{3})$ with two constant coefficients $C_1$ and $C_2$.
  We determine the $C_1$ and $C_2$ by the continuity of the potential and its derivative at the point of the transition, after expressing $J_{3/2}$ and $Y_{3/2}$ in terms of spherical Bessel functions.
The plot of the absolute value of Eq.~(\ref{pot}) is depicted in Fig.~\ref{potentialPlot}.
The $\Phi(x)$ remains constant as long as $x\ll 1$. 
At the horizon crossing, $\eta=\eta_\text{entry}$, the gravitational potential starts decaying as Eq.~(\ref{pot}) dictates: roughly as $x^{-3/2}$ during eKD and after the transition to the RD phase, as $x^{-2}$.

The reheating conformal time $\eta_\text{rh}$ is inversely proportional to the reheating temperature, $\eta_\text{rh}\propto T_\text{rh}^{-1}$.
This means that higher reheating temperature scenarios predict a faster reheating transition into RD. 
During eKD the growth of IGWs goes proportional to conformal time for a given wave mode $k$; i.e., $\Omega_{\textrm{IGW}}^{\textrm{(KD)}}\sim (k_\text{p}\eta)^2\overline{I^2_{\textrm{KD}}}
\propto\eta\propto a^2$.
This growth stalls once the transition is completed and radiation takes over. 
In the cases that $\eta_\text{entry}\ll \eta_\text{rh}$, we neglect the effect from RD era, because by that time the gravitational potential sourcing the tensor modes is negligibly small.
Numerically, in those cases we calculate the IGWs at the time of transition\footnote{After the transition, a sharp peak appears at $k={2\over\sqrt{3}}k_\text{p}$ superimposed on the spectral shape obtained at $\eta=\eta_\text{rh}$.
This corresponds to a logarithmic divergence manifest in the radiation domination as can be seen in Eq.~(\ref{radiation_TTF}).
We neglect such a resonance peak as it has no physical meaning, but it is rather a mathematical singularity associated to the delta power spectrum we used, Eq.~(\ref{deltaD})}.
Therefore, sudden eKD to RD scenarios  may show up in the experiments with an enhanced IGW spectrum.

\begin{figure}[h]
    \centering
    \includegraphics[width=8.2cm,height=5cm]{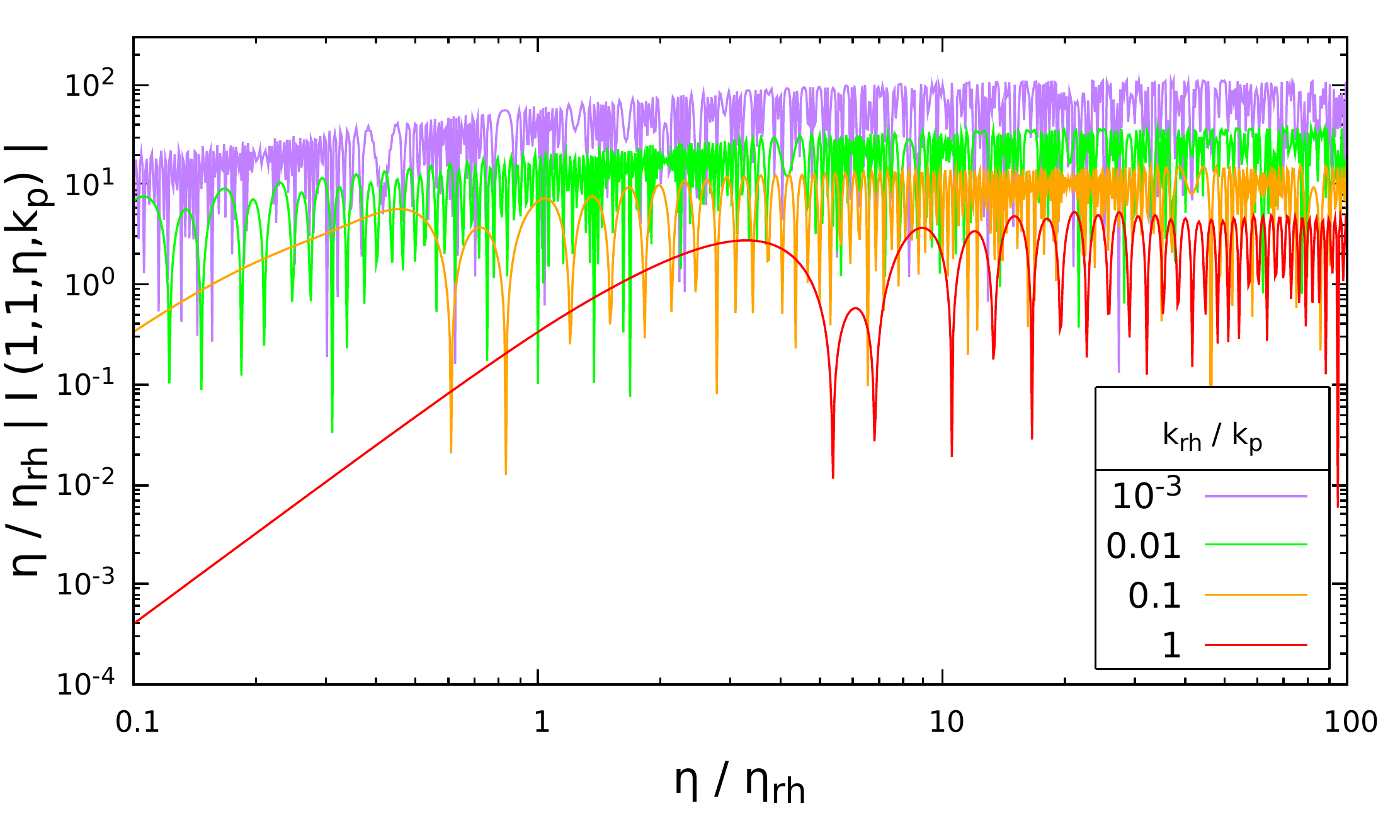}
    \caption{The behavior of the time-dependent piece of the IGW spectrum under a sudden eKD$\rightarrow$RD transition. The transition occurs when $\eta/\eta_\text{rh}=1$. Here we plot the combination $\left(\eta/\eta_\text{rh}\right)|I|$ for $k_\text{p}/k=1$ and for four choices of the ratio $k_\text{rh}/k_\text{p}$ corresponding to different colors in this figure. The growth is proportional to $\sqrt{\eta}$ during eKD. After the transition, this function always approaches an oscillatory state with a constant amplitude in the late times.}
    \label{eye}
\end{figure}

When there is a fast reheating, i.e.,  right after the horizon entry of the perturbation the universe transits into the radiation regime,
the growth, $\Omega_\text{IGW}\propto a^2$, of IGWs ceases early. 
However, the gravitational potential has not been suppressed significantly, and the contribution from RD era is important. 
In such a case of fast reheating we calculate the IGWs  at a conformal time $\eta_\text{c}=50\ \eta_\text{rh} \gg \eta_\text{rh}$,  thus taking into account the non-negligible contribution from the RD era.
In Fig.~\ref{eye}, we present the evolution of the time-dependent piece of the IGW  spectral shape with respect to the conformal time for the transition eKD$\rightarrow$RD.

The final formula for the density parameter of IGWs in the eKD$\rightarrow$RD transition is given by
\begin{eqnarray}
    \Omega_{\textrm{IGW}}^{\textrm{eKD}\rightarrow \textrm{RD}}(\eta_\text{c},k)={1\over6}A_{\mathcal{R}}^2\ \left({k_\text{p}\over\mathcal{H}(\eta_\text{c})}\right)^2\ \left[1-\left(k\over2 k_\text{p}\right)^2\right]^2
    \nonumber
    \\
    \times\ \overline{I_{\textrm{eKD}\rightarrow\textrm{RD}}^2\left({k_\text{p}\over k},{k_\text{p}\over k},\eta_\text{c},k\right)}\ \Theta\left(1-{k\over2 k_\text{p}}\right).\ \ \ \ \ \
\end{eqnarray}
Here the unit step function $\Theta$ is included by conservation of momentum, so that tensor modes with $k>2k_\text{p}$ are cut off. We remind the reader that we have assumed a delta-type power spectrum for the curvature perturbation. Later, in our results in Sec. \ref{SecResults} and conclusions in Sec. \ref{secConclusions}, we estimate the amplitude of the $\Omega_\text{IGW}$ for a broad curvature power spectrum, of Gaussian type, at the peak frequency.

Finally, let us mention that the existence of an early kination phase implies that the energy density of the produced IGW gets enhanced and might backreact on the geometry. This backreaction must be limited in order that the energy density of the IGWs during the BBN does not increase the expansion rate to a disturbing level for the observed abundances of the relic nuclei. In order to comply with this constraint we find out a new bound on the reheating temperature that must be satisfied,
\begin{align}
   {T_\text{rh}}\gtrsim  10^{7}\,{\text{GeV}} \,
    A_{\cal R}^{3/2}\left(\frac{M_\text{}/\gamma}{10^{20}\text{g}} \right)^{-1/2}.
\end{align}
This is a rough, conservative bound that connects horizon mass $M/\gamma$ and the $A_{\cal R}$  with the $T_\text{rh}$, and guarantees the success of the BBN predictions. 
Further discussion and derivation details can be found in Appendix~\ref{TmaxKin}.

\section{Primordial black holes and the associated Gravitational Wave signal} \label{SecPBH}

PBHs form from the collapse of large-amplitude primordial inhomogeneities \cite{Carr:1974nx,Meszaros:1974tb, Carr:1975qj, Lindley:1980bu}. An inhomogeneity decouples from the background expansion 
if the power spectrum ${\cal P_R}(k)$ is enhanced at a scale  $k^{-1}$, characteristic of the PBH mass. During RD, typical values for the ${\cal P_R}(k)\sim {\cal O}(10^{-2})$ are required.
If PBHs have mass $M< 10^{15}$ g evaporates at time scales less than the age of the universe, whereas PBHs with $M > 10^{15}$ g survive till today being  dynamically cold component of the dark matter in galactic structures.

The PBH dark matter scenario is constrained in a wide range for the mass parameter $M$ by several observational experiments, labeled with  extra galacric gamma-ray background (EGB), V, GC, HSC, O, EM, LIGO/VIRGO, GW2, supernova (SN), wide binaries (WB), E, x-ray binaries (XB), Planck (PA) in Fig.~\ref{FigPBH}.
In the present universe, 
 the abundance of light PBH is constrained from the extra galactic gamma-ray background \cite{Page:1976wx, MacGibbon:1991vc, Carr:1998fw, Barrau:2003nj,  Carr:2016hva}, as well as from positron constraints (V) \cite{Dasgupta:2019cae}.
Black holes of mass above $10^{17}$g are subject to gravitational lensing constraints  \cite{Barnacka:2012bm, Tisserand:2006zx, Niikura:2017zjd}, given by Subaru (HSC), Ogle (O), EROS (E), and MACHO (M), microlensing of SN and others.
The CMB anisotropies measured by PA constrains the PBH with mass above $10^{33}$g 
\cite{Ricotti:2007au, Carr:2016drx, Clesse:2016vqa, Bird:2016dcv, Poulin:2017bwe, Serpico:2020ehh}. At the large mass region there are also constraints from accretion limits in X-ray and radio observations \cite{Gaggero:2016dpq} and XB \cite{{Carr:2020gox}}. There are also dynamical limits from disruption of  WB, survival of star clusters in Eridanus II.
Second order tensor perturbations (GW2) generated by scalar perturbations already constrain the abundance of PBH masses approximately in the range $10^{30}- 10^{33}$ g. Also, the expected GWs generated by PBH binaries with mass ${\cal O}(1-10)M_\odot$ coalescing at the present epoch is constrained by LIGO/Virgo \cite{Sasaki:2016jop, Eroshenko:2016hmn, Raidal:2017mfl} as well as compact binary systems with component masses in the range  $0.2-1 M_\odot$ \cite{Abbott:2018oah} shown in Fig. \ref{FigPBH}.
For a recent update on the PBH constraints see Ref. \cite{Carr:2020gox}.

\subsection{PBH abundance for general EOS}

The present relic energy density parameter of primordial black holes with mass $M$ produced at the cosmic time $t$ is
\begin{equation} 
\label{fPBH}
\Omega_\text{PBH}(M)={\Omega_\text{m}}
\gamma \beta(M) \left(\frac{M_\text{rh}}{M/\gamma}\right)^
\frac{2w}{1+w}
\left(\frac{M_\text{eq}}{M_\text{rh}}\right)^{1/2} 
\tilde{g}(g_*) 
\end{equation}
where $\tilde{g}(g_*)=2^{1/4} 
 (g_*(t)/g_*(t_\text{rh}))^{-s/4}(g_*(t_\text{rh})/g_*(t_\text{eq}))^{-1/4}$ 
and $g_*$ the thermalized  degrees of freedom.
The parameter $s$ is equal to 1 for $t>t_\text{rh}$ and 0 for $t<t_\text{rh}$.
$\Omega_\text{m}$ is the total matter density parameter today, $M_\text{eq}$ is the horizon mass at the moment of matter radiation equality, 
and $M_\text{rh}$ the horizon mass at the moment of reheating.

The $\beta(M)$ is 
the mass fraction of the universe with horizon mass $M/\gamma$ that collapsed and formed PBHs; it can be interpreted as the black hole formation probability.
Assuming Gaussian statistics, for a spherically symmetric region it is 
\begin{equation} \label{brad}
\beta_\text{}(M)=\int_{\delta_c}d\delta\frac{1}{\sqrt{2\pi\sigma^2(M)}} e^{-\frac{\delta^2}{2\sigma^2(M)}}\,,
\end{equation}
where $\sigma(M)$ is the variance of the density  perturbations and $\delta$ the density contrast.
The PBH abundance has an exponential sensitivity to the variance of the perturbations $\sigma(M)$ and  the threshold value $\delta_c$, which is $w$ dependent. In this work, we assign values to $\delta_c$ following the findings of Ref. \cite{Harada:2013epa}. 

\begin{figure}[h]
    \centering
    \includegraphics[width=8.5cm,height=5cm]{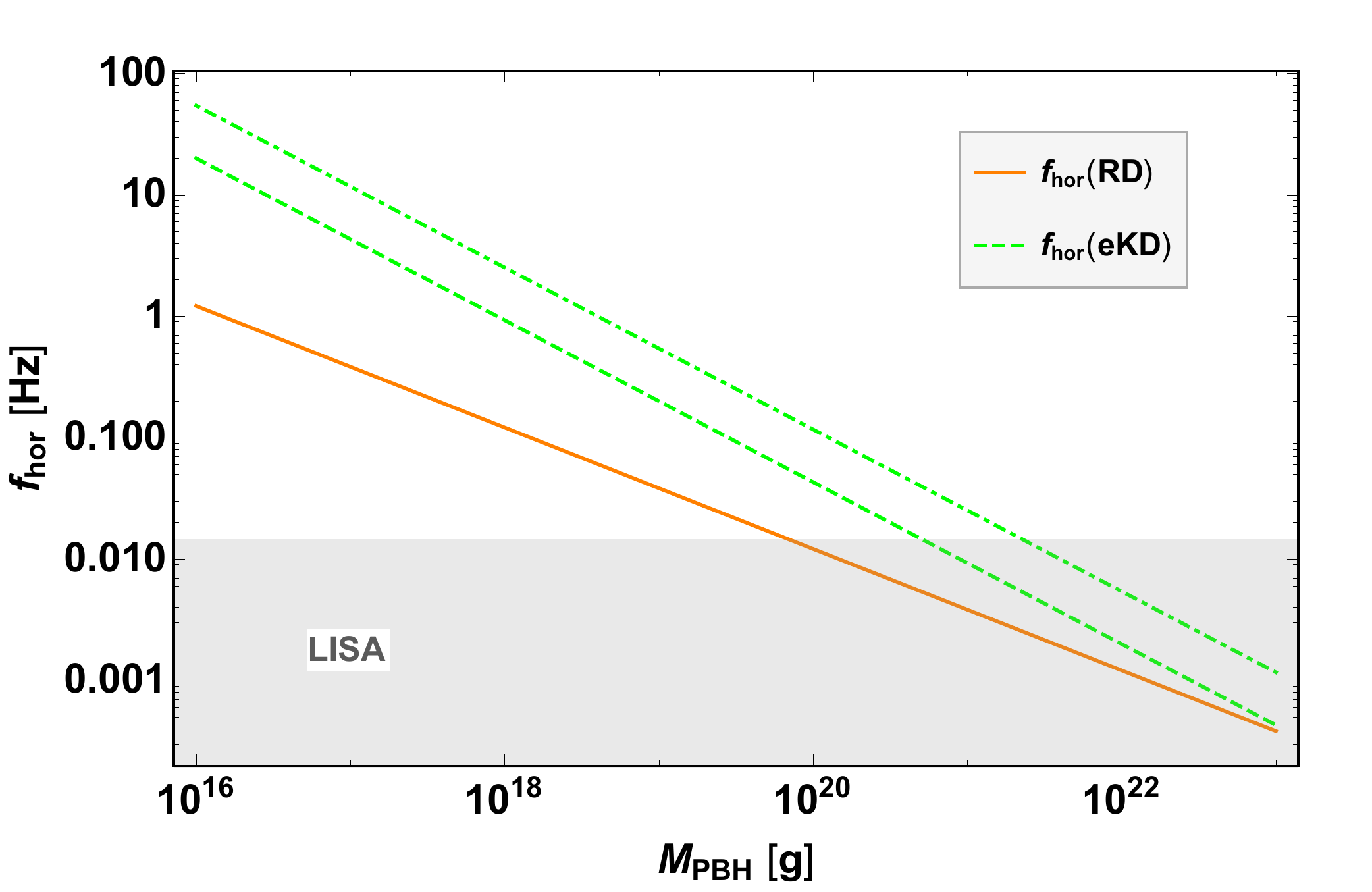}
    \caption{The figure depicts the relation between the PBH mass $M_\text{PBH}$ and the corresponding frequency of horizon with size $k^{-1}=(2\pi f_\text{hor})^{-1}$ and mass $M_\text{PBH}/\gamma$ for two equation of states  $w=1/3$ (orange line) and $w=1$ (green lines); see Eq.~(\ref{kMgen}). 
    The reheating temperature is $T_\text{rh}=10^4$ GeV ($5\times 10^2$ GeV) for the dashed (dot-dashed) lines.
    The gray strip highlights the frequency band of LISA optimal sensitivity.}
    \label{Fig.fMplot}
\end{figure}

The horizon mass increases with a different rate for different expansion rates.
Using the relation $f=k/(2 \pi)$ we find the frequency-PBH mass correspondence for general  equation of state $w$ and reheating temperature $T_\text{rh}$,
 \begin{align} \label{kMgen}
f_\text{hor} (M,  T_\text{rh}, &  w)  \simeq 2.7 \times 10^{2} \text{Hz} \left( \frac{T_\text{rh}}{10^{10}\text{GeV}} \right)^{\frac{1-3w}{3(1+w)}}  \nonumber \\
& \left(\frac{M/\gamma}{10^{12} \text{g}} \right)^{-\frac{3w+1}{3(1+w)}} \left(\frac{g_*}{106.75}\right)^{-\frac{2w}{6(1+w)}}\,.
\end{align}
In the above relation, \footnote{For the case of kination domination the minor correction of replacing $T_\text{rh}$ by $2^{1/4} T_\text{rh}$ should be considered if there is equipartition between radiation and scalar field  energy density at the time of reheating.} we have assumed a one-to-one correspondence between $k$ (or$f$) and the PBH mass $M$.
This is true for the approximation of a monochromatic PBH mass spectrum, which is practically the case in many models. We note that the PBH mass distribution peaks at a value $M(k)$ that is in  a slight offset with the value $M(k_\text{p})$, as our following numerical analysis shows. In the $k$-space the peak of the $\sigma(k)$ is about at $0.7 k_\text{p}$ \cite{Wang:2019kaf}. The $f (M,  T_\text{rh},  w)$ relation is depicted in Fig.~\ref{Fig.fMplot}.
The $f_\text{hor}(M)$ frequency and $f_\text{p,IGW}$ frequency, where the IGW spectrum maximizes, have similar size but do not coincide. 
 
 Therefore, searching for GWs with frequency $f_\text{IGW}$, one can probe PBH scenarios and the reheating temperature of the universe.

\subsection{GW detectors and PBH scenarios}
\label{sub:GWDs}

Currently, there is a network of operating ground-based GW detectors  that focus on the high and low frequency regime, roughly at $10^2$ and $10^{-9}$ Hz, of the GW spectrum.
Additionally, there are several designed and proposed experiments sensitive enough to detect or constrain the stochastic GW background at a large range of frequencies. 
In this subsection, we outline the GW experiments labeled in Fig.~\ref{FigIGW} that can test the predictions of the models described in this work.

The operating ground-based GW detector system is the advanced Laser Interferometer Gravitational wave Observatory (aLIGO). In Fig.~\ref{FigIGW} we show for reference O1 and O5 which correspond to the first observing run and the design sensitivity respectively \cite{Aasi:2013wya}.
The future ground-based laser third generation interferometers is the Einstein Telescope (ET) \cite{Sathyaprakash:2012jk}, that probes the range of frequencies near 100 Hz.
The space-based proposed experiments are DECIGO \cite{Seto:2001qf, Sato:2017dkf} and BBO \cite{Crowder:2005nr}, that will probe in the decihertz frequency window.  
In the millihertz band of the spectrum there is the scheduled LISA, \cite{Audley:2017drz}, and proposed space-based experiments such as the TianQin, \cite{Luo:2015ght}. Pulsar Timing Arrays, with the international PTA (IPTA) \cite{Hobbs:2009yy}, and the planned Square Kilometer Array (SKA) \cite{Janssen:2014dka} are projects that probe and already pose constrains on the nanohertz regime.

Fitting functions regarding the sensitivity curves for LISA, aLIGO and ET are provided by \cite{Sathyaprakash:2009xs};  for DECIGO and BBO, we consulted \cite{Yagi:2011wg} and adjusted the sensitivities of these experiments to current values and expectations; 
for TianQin, we used the fitting function provided in Ref. \cite{Huang:2020rjf}.

IGWs can constrain a large window of the PBHs masses. 
Currently, there are indirect constraints from the  pulsar timing array experiments on IGWs associated with the formation of relatively massive PBHs at the epoch of horizon entry.
Notably, a very severe constraint exists, $\beta(M) \lesssim 10^{-52}$, in the solar mass range, 
from pulsar timing data \cite{Saito:2008jc, Chen:2019xse}. 
We note that Ref.~\cite{Cai:2019elf} pointed out that the tension with PTA constraints can be relieved if the perturbations are locally non-Gaussian. 

\section{Results} 
\label{SecResults}

In this section we outline our findings. We assume ${\cal P_R}(k)$ amplitudes and wave numbers having as a ruler the PBH abundance, and then we calculate IGWs aiming at constraining early universe cosmological scenarios and inflationary models.
 
We consider ${\cal P_R}(k)$ peaks generated by the $\alpha$-attractors \cite{Dalianis:2018frf} and nonminimal derivative coupling Horndeski inflation  \cite{Dalianis:2019vit},
as well as, irrespectively of the inflationary theory, Gaussian and Dirac $\delta$-type distributions.
The peak choice of the $\mathcal{P}_{\mathcal{R}}$ is spanning a wide range of frequencies in the GW spectrum; see Fig.~\ref{Figps}.
We compare the resulting IGW shape with the sensitivity of current, planned and proposed GW experiments.
We choose amplitudes and shapes for the  ${\cal P_R}(k)$ that induce GWs and PBH abundances of cosmologically significant amount. 
We always make sure that our input parameters respect the current constraints for the PBH abundance and the GWs.
We also take care the position and the width of the ${\cal P_R}(k)$ peak to respect constraints coming from the Hawking radiation \cite{Dalianis:2018ymb}.

We assemble and present our results considering  first ${\cal P_R}(k)$  that generate a sizable amount of both PBHs and IGWs and secondly ${\cal P_R}(k)$ that generate only IGWs with the PBH counterpart being either negligibly small, $\Omega_\text{PBH} \lesssim 10^{-10}$, or promptly evaporating.

\subsection{IGWs in scenarios with dominant and subdominant PBH dark matter component}

\subsubsection{Significant PBH abundance}
In the mass window $M_\text{}= 10^{17}-5\times 10^{22}$ g,
the PBH abundance can reach its maximum value,\footnote{In that range there are  feeble constraints coming from white dwarfs and neutron stars \cite{Capela:2012jz,Capela:2013yf, Graham:2015apa, Brandt:2016aco} 
but these constraints, together with others in the same mass range such as femtolensing and picolensing, are seen as insecure \cite{Katz:2018zrn, Montero-Camacho:2019jte, Carr:2020gox}.} $\Omega_\text{PBH}/\Omega_\text{DM}=1$; see Ref. \cite{Carr:2020gox}. 
If PBHs constitute a significant fraction of the total dark matter then an IGW counterpart must exist that can be tested by 
LISA and by 
 proposed experiments such as BBO and DECIGO.

Another motivated PBH mass range is $M={\cal O}(1-100) M_\odot$ where LIGO detected several coalescence events the last years with the most recently published the intriguing event \cite{Abbott:2020khf}. For that PBH mass range the IGWs are expected to produce GWs with nanohertz frequency  detectable by PTA experiments \cite{Chen:2019xse}.  

The exact PBH mass parameters that we choose can be found in Tables \ref{tabRad} and \ref{tabKin}
and in  Fig. \ref{FigPBH}. In the tables,  the order of magnitude value for PBH abundance  is quoted  due to the exponential sensitivity to the  model-dependent threshold value $\delta_c$. 
\begin{figure*}[t]
    \centering
     \includegraphics[width=8.5cm,height=5cm]{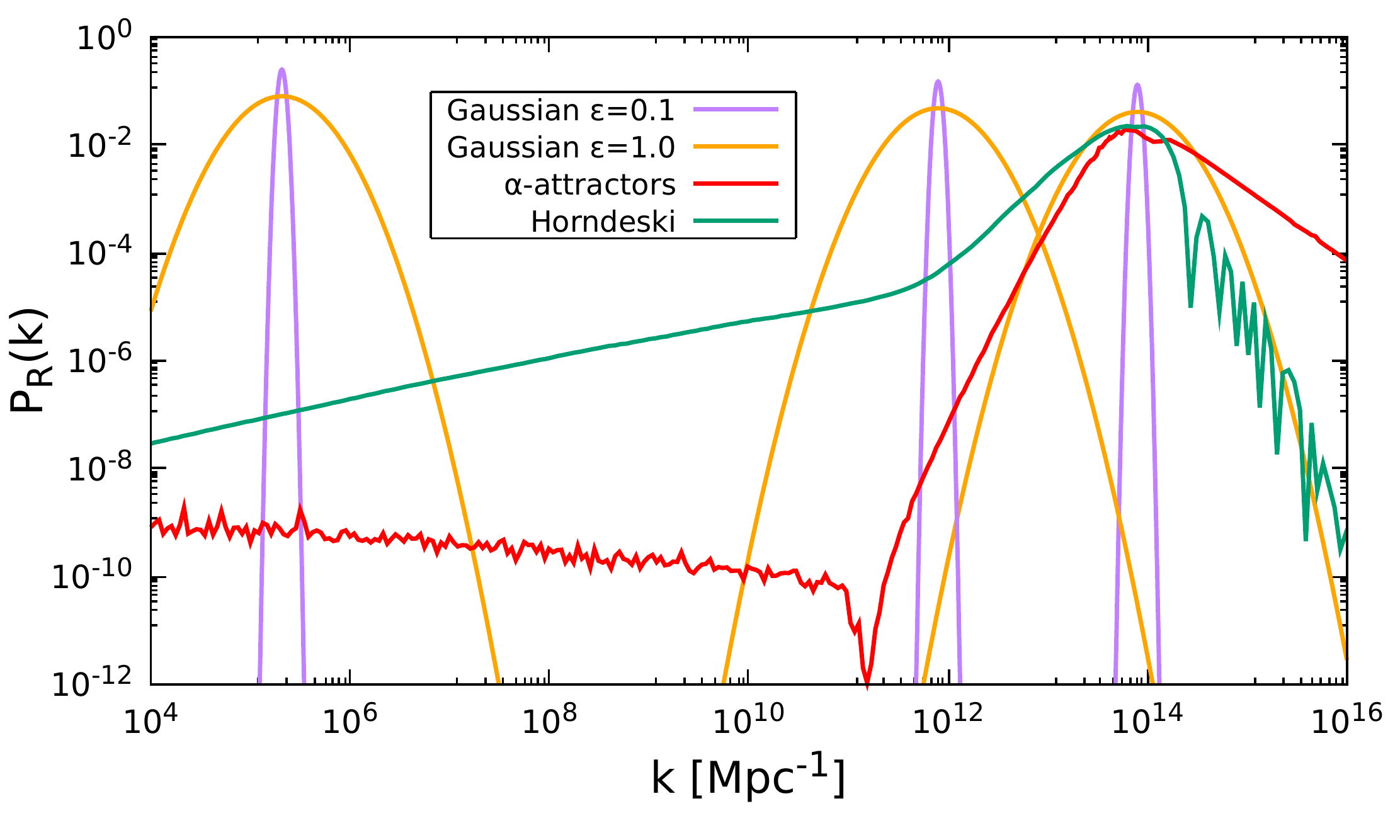}
     \includegraphics[width=8.5cm,height=5cm]{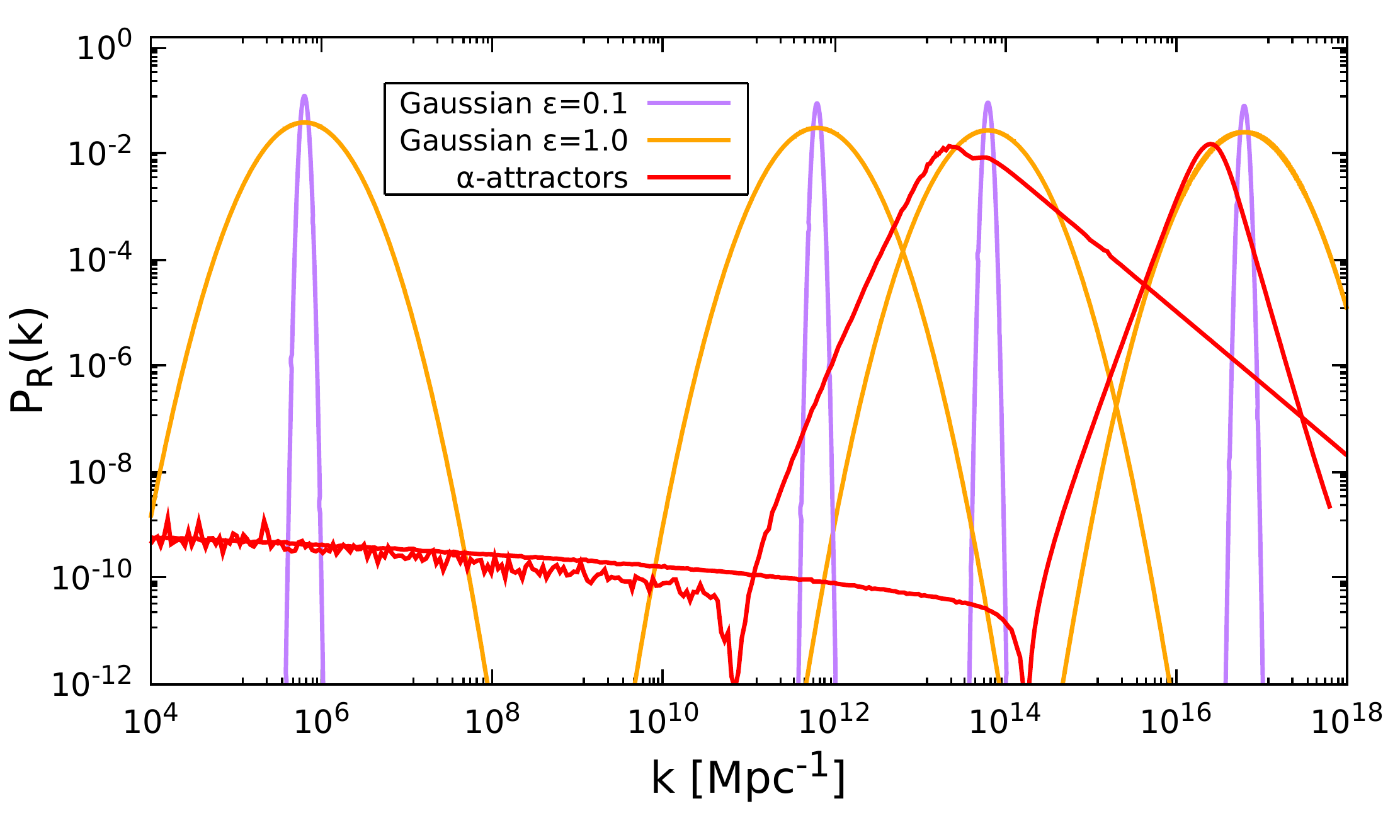}   
    \caption{The power spectra of curvature perturbations sourcing IGWs during RD. We consider a medium and a narrow Gaussian shapes for ${\cal P_R}(k)$ as well as ${\cal P_R}(k)$ produced by  $\alpha$-attractors (red curves) and Horndeski nonminimal derivative coupling inflation models (green curve). 
    The scaling of each ${\cal P_R}(k)$ slope is given in Table \ref{powerTable}.
    Oscillations in the red curves correspond to numerical effects.
    \textit{Left panel:} ${\cal P_R}(k)$ that generates significant PBH abundances, $\Omega_\text{PBH}\sim {\cal O} (0.01-1)$. \textit{Right panel:} ${\cal P_R}(k)$ that generates negligible PBH abundances, $\Omega_\text{PBH}\lesssim 10^{-10}$.}
    \label{Figps}
\end{figure*}

\begin{figure*}[t]
    \centering
    \includegraphics[width=8cm,height=4.7cm]{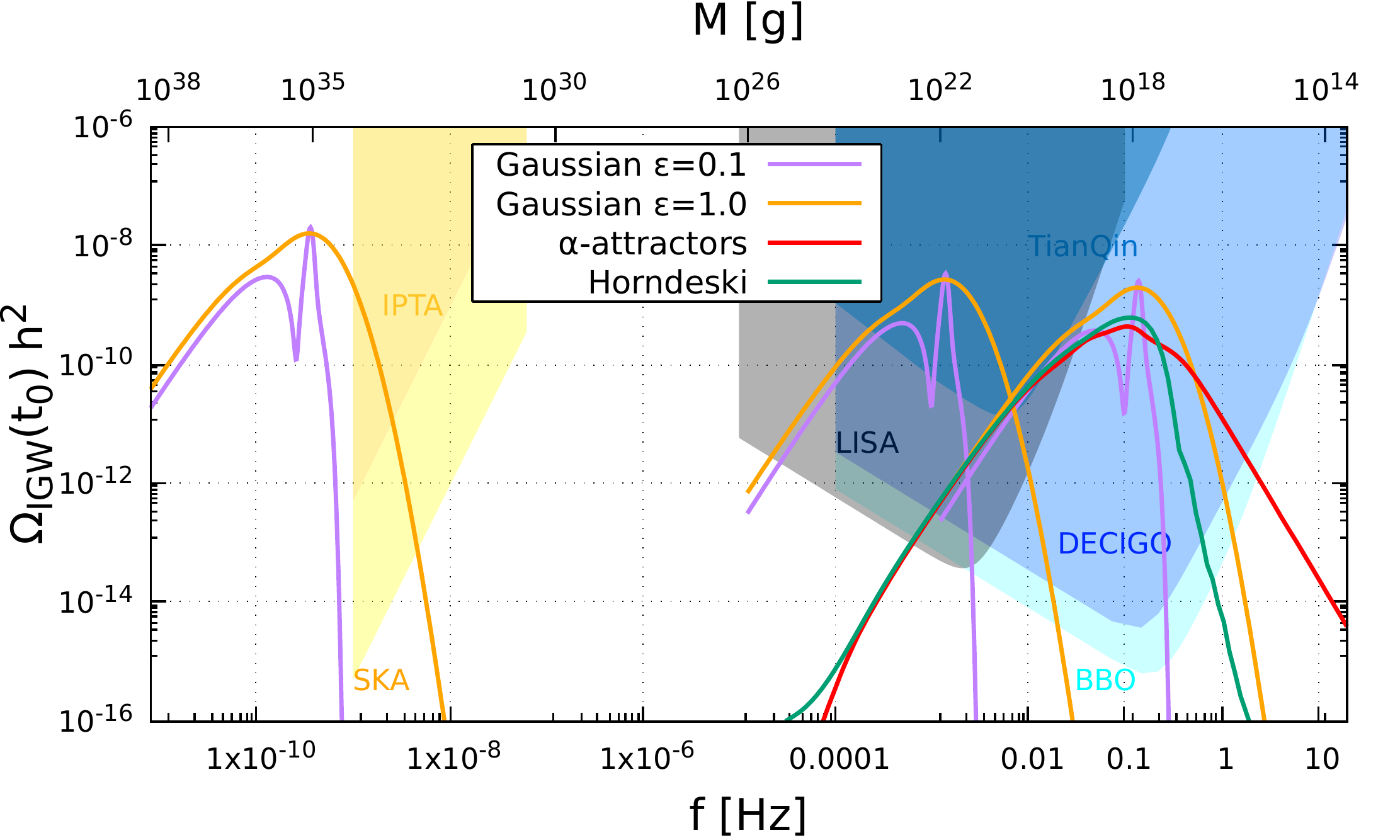}
    \includegraphics[width=8cm,height=4.7cm]{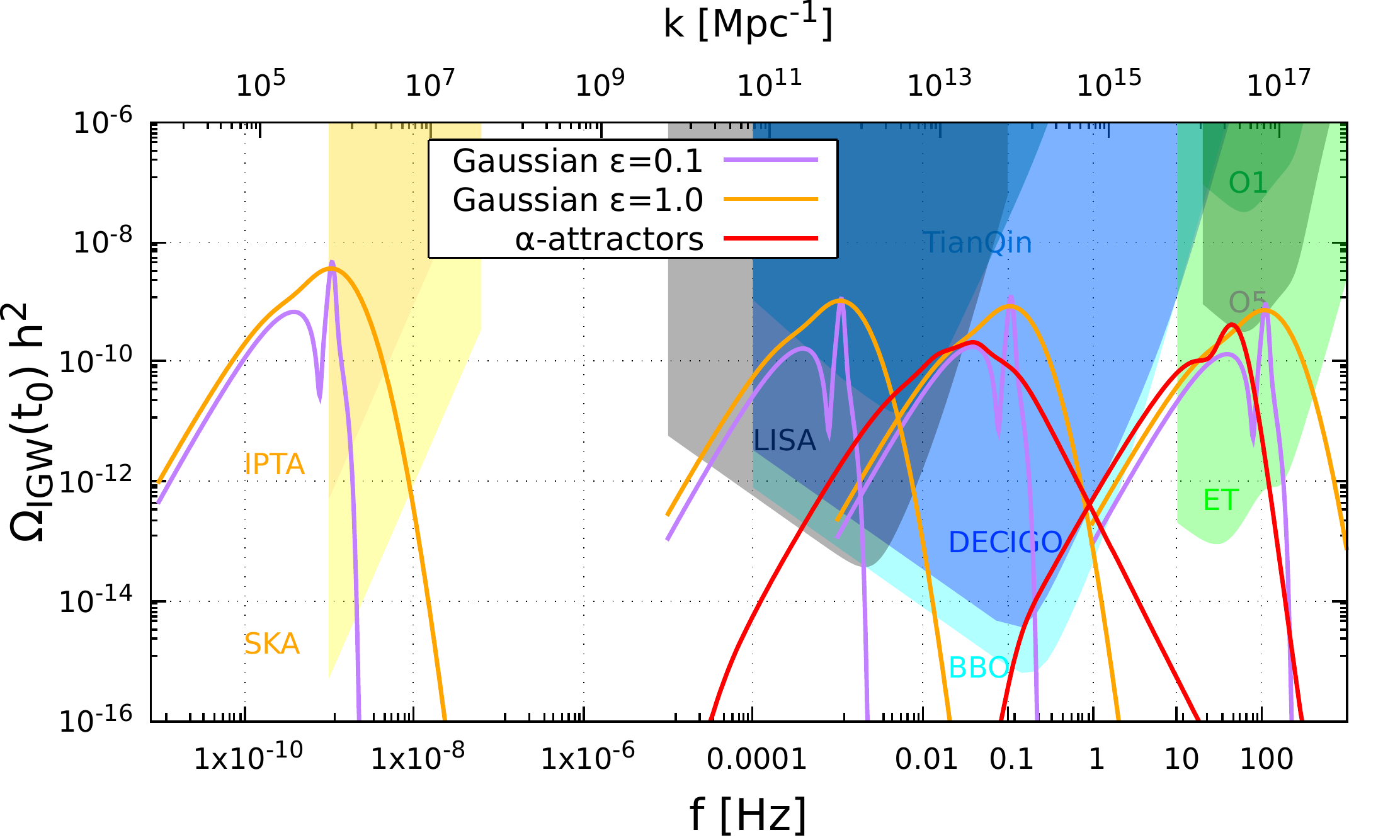}
    \includegraphics[width=8cm,height=4.7cm]{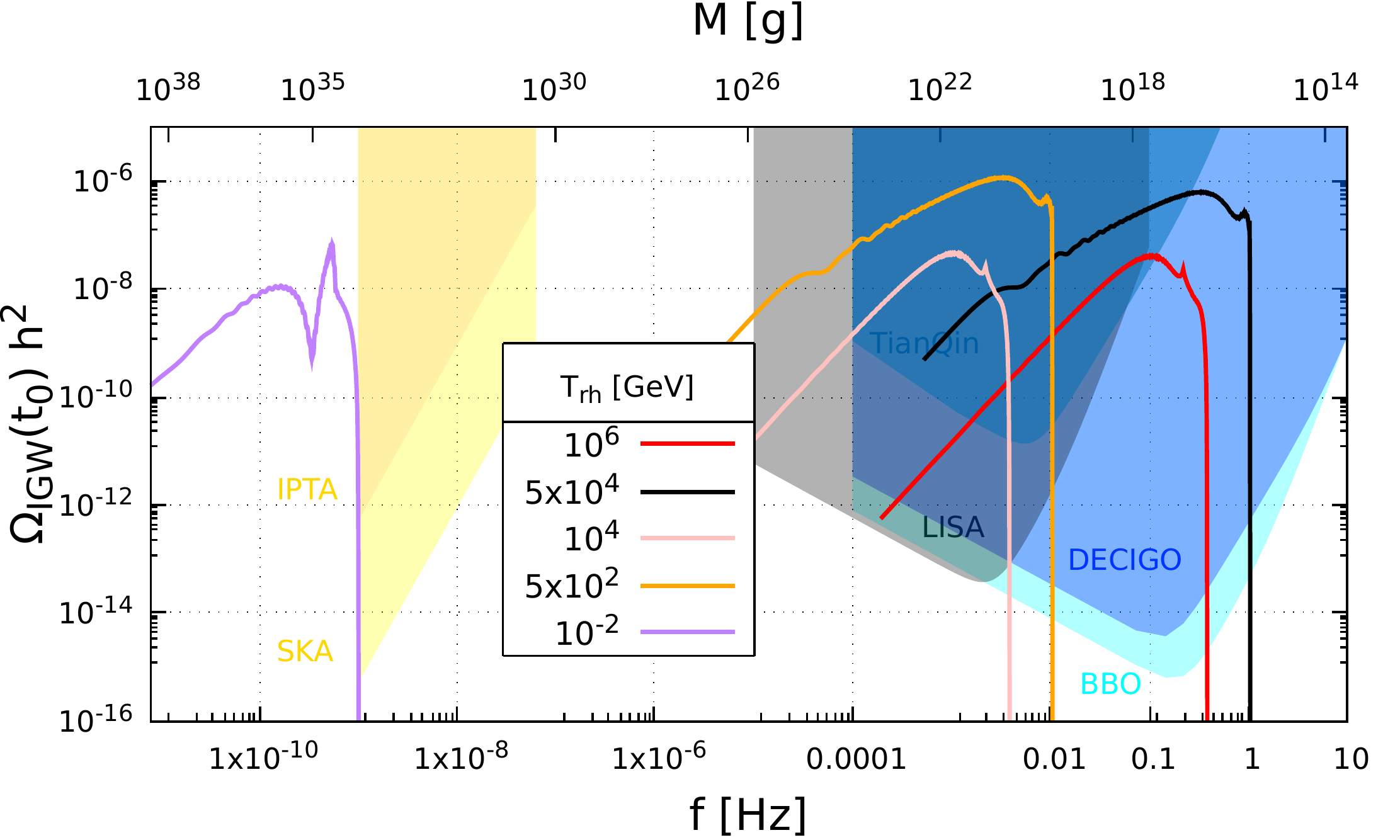}
      \includegraphics[width=8cm,height=4.7cm]{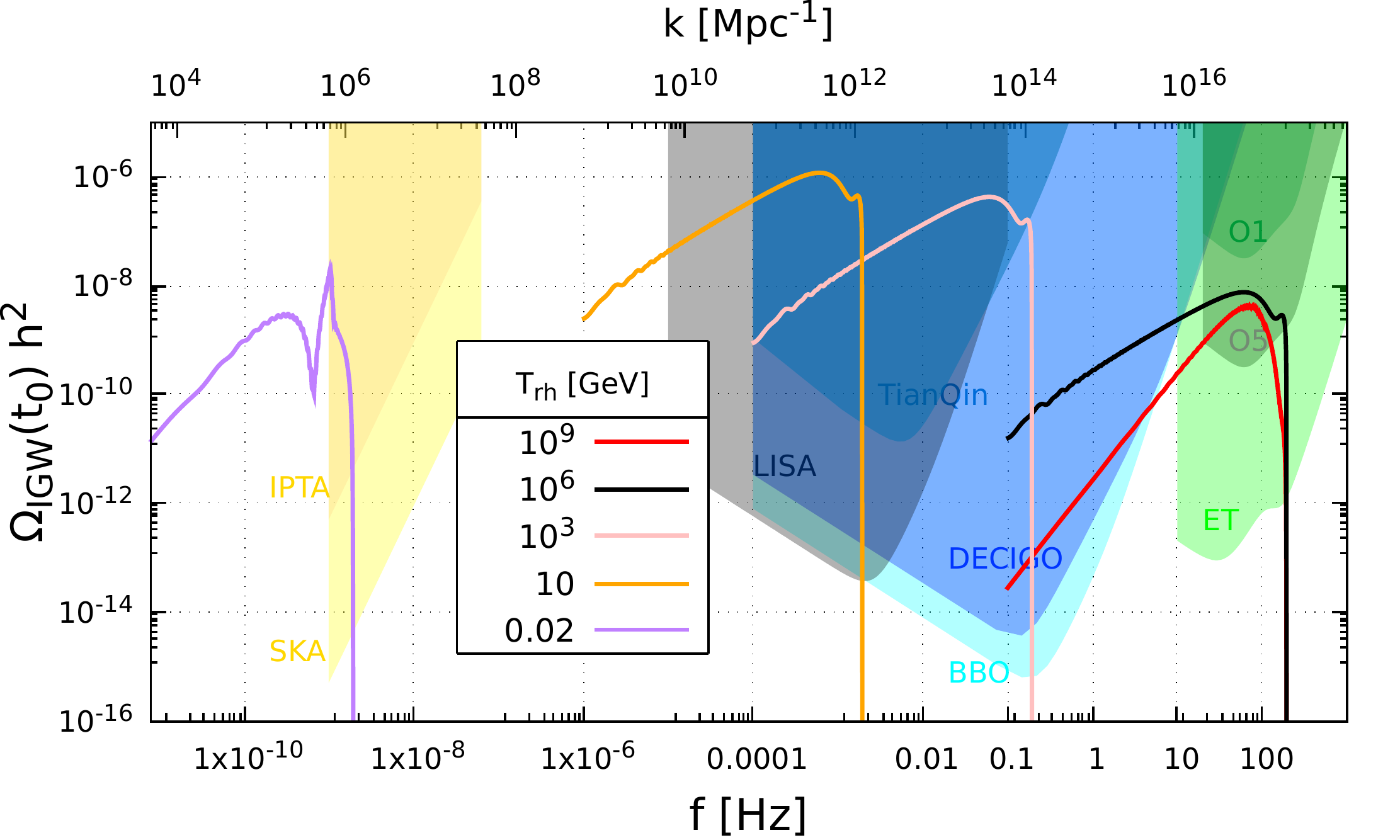}
    \caption{The IGW spectral shapes for two early universe scenarios, the RD ({\it top panels}) and the eKD$\rightarrow$RD ({\it bottom panels}) sourced by ${\cal P_R}(k)$ depicted in Fig.$~\ref{Figps}$. In the {\it left panels} we consider the scenario  of abundant PBH production  and in the {\it right panels}  negligible PBH abundances. In the background of the IGW spectral curves the sensitivity curves of current/planned/proposed GW detectors are shown as described in the subsection~\ref{sub:GWDs}. The parameters for the ${\cal P_R}(k)$ and the derived values for the IGWs are listed in the Tables \ref{tabRad}-\ref{tabKinf}.}
    \label{FigIGW}
\end{figure*}
\begin{figure*}[t]
    \centering
    \includegraphics[width=8cm,height=4.7cm]{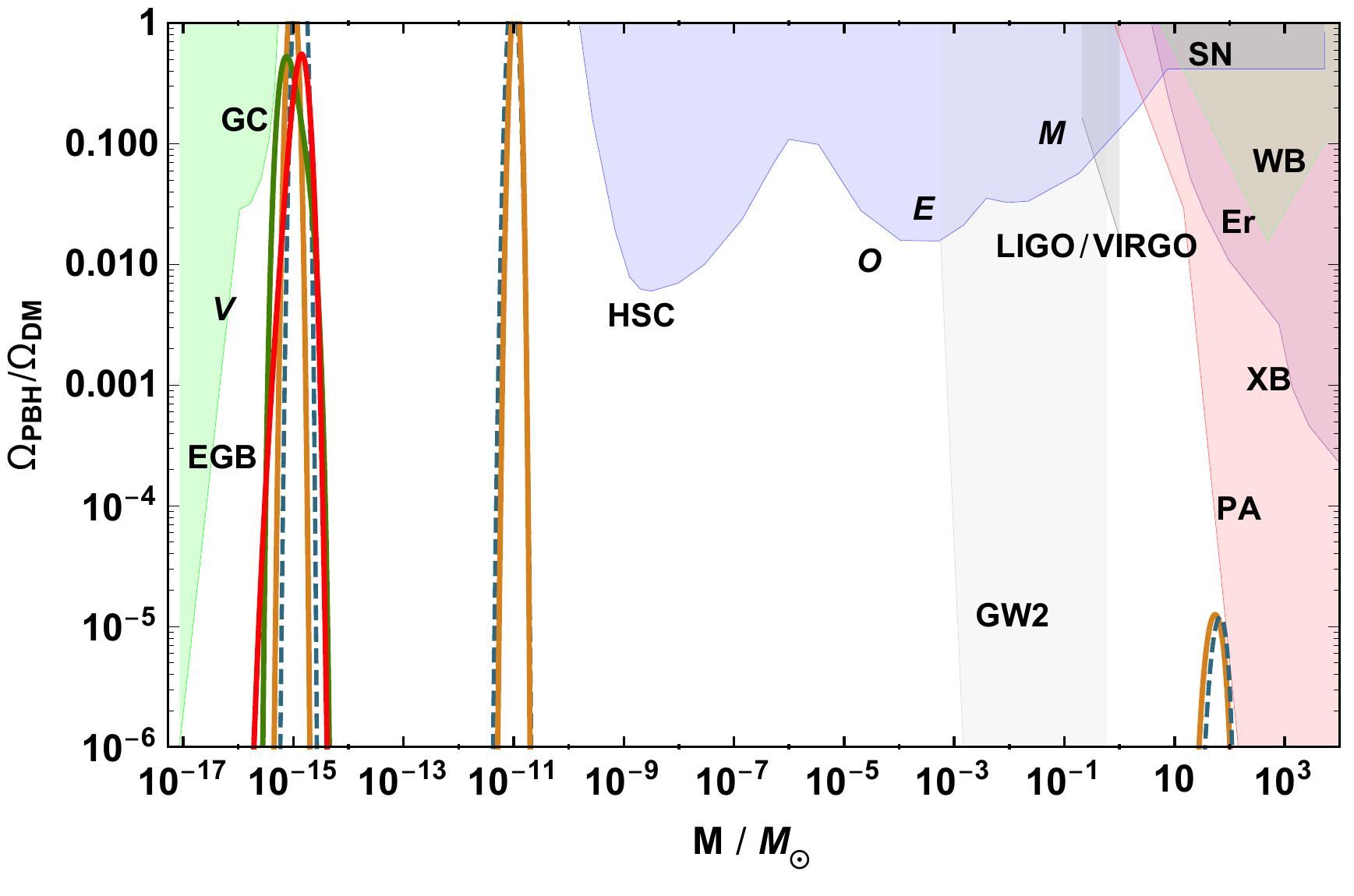}
    \includegraphics[width=8cm,height=4.7cm]{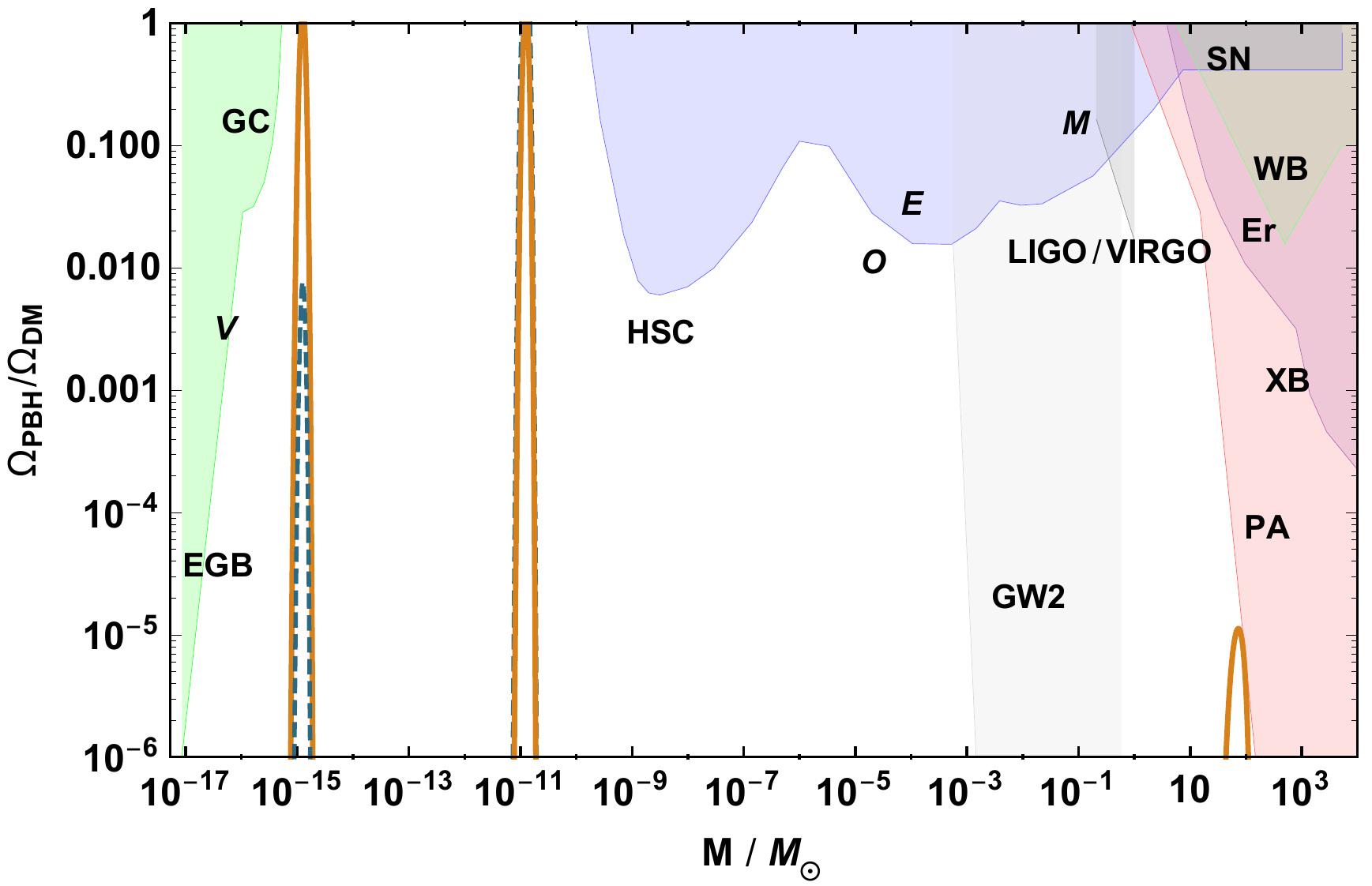}
    \caption{The fractional abundance of the PBHs produced during  RD ({\it left panel}) and eKD  ({\it right panel}) for our ${\cal P_R}(k)$ scenarios. The colored upper bounds are the present experimental constraints as described in Sec.~\ref{SecPBH}.  In the left panel, the red curve  corresponds to the $\alpha$-attractor inflationary model, the green to the Horndeski nonminimal derivative coupling model, the orange to a Gaussian of width $\epsilon= 1$, and the dashed to a Gaussian of width  $\epsilon=0.1$. The ${\cal P_R}(k)$ for each curve is depicted in Fig.\ref{Figps}. In the right panel, we take a $\delta$-distribution for the $\cal P_R$;  also, for the orange curves the reheating temperature is higher than for the dashed curves. 
    Detailed values can be found in Tables \ref{tabRad} and \ref{tabKin}, respectively; the associated IGWs are depicted in Fig. \ref{FigIGW}. $M_\odot$ stands for the sun mass, roughly $2\times10^{33}$ g.}    \label{FigPBH}
\end{figure*}
\subsubsection{Negligible PBH abundance}

Contrary to the exponential sensitivity of the $\Omega_\text{PBH}$ on the $A_{\cal R}$, the sensitivity of the IGWs amplitude on the $A_{\cal R}$ is of a power-law type. Therefore, a minute decrease in the ${\cal P_R}(k)$ amplitude can nearly disappear the PBHs abundance while decrease the IGW amplitude only by a minor amount.
This is rather fascinating because primordial density perturbations washed out completely by the RD era have left a relic GW behind that reveals their presence. 

In addition, there are PBH masses that evaporate promptly in the early universe. These PBHs do not survive until today and, unless a stable remnant is left behind, the corresponding $\Omega_\text{PBH}$ is zero and cannot account for the dark matter in the galaxies. Nevertheless, the associated IGWs might be strong enough to be  detectable. Such a possibility is rather interesting  
because the PBH remnants can constitute a significant fraction of the total $\Omega_\text{DM}$ and, moreover, the inflationary scenarios that predict mini-PBHs  can have a spectral index $n_s$ in full accordance with the Planck-2018 data \cite{Dalianis:2019asr},
see Eq.~(\ref{Vrun}) for the $\alpha$-attractors potential that we use. 

Our results for zero or negligible PBH abundance are listed in Tables \ref{tabRadf} and \ref{tabKinf}. We note that, instead of $\Omega_\text{PBH}$, we write the $\beta(M)$ values, see Eq.~(\ref{brad}), since PBH with mass $M\lesssim 10^{15}$ g has already evaporated.

\subsection{Cosmological eras}

\subsubsection{Radiation}

The spectral shapes of the IGWs produced during the radiation era are depicted in the upper panels of Fig. \ref{FigIGW}. 
For narrow Gaussian ${\cal P_R}(k)$ peaks, $\epsilon=0.1$,  the spectral shape obtained has  a double peak structure with a sharp peak at $f\simeq 2f_\text{p}/\sqrt{3}$, resembling the IGW spectrum generated by a $\delta$-distribution for ${\cal P_R}(k)$. For a broader ${\cal P_R}(k)$ peak a single wide peak forms in the IGWs at approximately $2f_\textrm{p}/\sqrt{3}$, see top panels of Fig.~\ref{FigIGW}.

The $\alpha$-attractor inflationary model, described by the potential (\ref{Vattrpol}),  generates an enhanced ${\cal P_R}(k)$ as depicted in Fig.~\ref{Figps}. The predicted IGW spectrum resembles more that of the broad Gaussian, $\epsilon=1$, with a peak shifted towards $k_\text{p}$, i.e. $f_\text{p,IGW}\simeq f_\text{p}$. The general nonminimal derivative coupling model, see Eqs.~(\ref{Lhorn}),~(\ref{Higgs}) and~(\ref{gnmdc}), features an IGW spectrum that resembles the broad Gaussian; hence it also resembles the $\alpha$-attractor model around the peak. However, the IGW spectra of the two inflationary models deviate at large frequencies following the different scaling at the high-$k$ tale of the ${\cal P_R}(k)$, as the inspection of Figs.~\ref{Figps} and~\ref{FigIGW} shows.

For the case the $A_{\cal R}$ amplitude is decreased to a level that the PBH abundance becomes negligible, $\Omega_\text{PBH} \lesssim 10^{-10}$,  the amplitude of the IGWs decreases a little and the spectral shape characteristics remain unchanged. For the case of evaporating mini-PBHs, we examine a different $\alpha$-attractors model, described by the potential (\ref{Vrun}). This model generates a ${\cal P_R}(k)$ peak that is relatively narrow. The spectrum of the IGWs is found to feature a double peak structure, manifest in  the Figs.~\ref{Figps} and~\ref{FigIGW}.

Regarding the energy density, 
we find that the growth of the IGWs amplitude occurs rapidly, for a period $\eta_\text{c}\sim O(10)k_{\text{p}}^{-1}$, and then the growth ceases. 
The growth is realized from the time of entry until the potential decreases significantly, since $\Phi(x)$ oscillates with a $x^{-2}$ decaying amplitude in subhorizon scales. Afterward the source term becomes negligible and the GW propagates freely.
At that time $\eta_c$
the $\Omega_\text{IGW}$ produced by a Gaussian ${\cal P_R}(k)$ peak
with width $\epsilon$
and at the frequency $f_\text{p}$ is
\begin{equation}
    \Omega^{(\text{RD})}_\text{IGW}(\eta_\text{c}, f_\text{p}) \sim \left( A_{\cal R} \, \epsilon \sqrt{\pi}\, \right)^2.
\end{equation}
A similar relation has been found following analytic steps in Ref. \cite{Pi:2020otn}. 
We further comment on the scaling of the IGW spectrum, at scales beyond the peak, in the next subsection. Note that in Tables \ref{tabRad}-\ref{tabKinf} the present total energy density value, $\Omega_{\textrm{IGW}}(t_0)=\int d\ln f\ \Omega_{\textrm{IGW}}(t_0,f)$,  is listed.

\subsubsection{Early kination}

The IGWs produced during the early kination era and sourced by $\delta$-distributions for  ${\cal P_R}(k)$ are  depicted in the lower panels of Fig.~\ref{FigIGW}.
The increase of the IGW amplitude is depicted in Fig. \ref{eye}.
What is of interest is that the shape of the IGW spectrum depends on the reheating temperature, always assuming a sudden change from eKD to RD at the temperature $T_\text{rh}$.

If the $\delta$ peak enters well before the 
reheating moment, $\eta_\text{entry} \ll \eta_\text{rh}$, the induced tensor power spectrum has a distinctive shape, characteristic of the  kination stage. It is distinctive because the scalar perturbations decay as $x^{3/2}$ experiencing a maximal pressure. 
No sharp peak in the IGWs appears even for the monochromatic ${\cal P_R}(k)$. This feature has been also noted in  \cite{Domenech:2019quo}.
Additionally, the energy density of the IGWs gets enhanced  due to the redshift of the stiffer background.  We call this transition "slow eKD$\rightarrow$ RD."

If, on the other hand, the reheating happens fast after the horizon entry of the $\delta$-peak, the scalar perturbations experience  the maximal pressure of the stiff fluid only for a  while. It is actually the radiation phase that mostly forms the IGW spectrum. 
 We call this transition "fast eKD$\rightarrow$ RD".
This is readily seen at the IGW spectrum in the PTA frequency range, as depicted in Fig. \ref{FigIGW}, where the BBN bound for the $T_\text{rh}$ limits the duration of the kination era. The parameters chosen for this case are $k_\text{p}\eta_\text{rh}\approx1.5$ which means that transition essentially occurs at the moment of entry of the curvature perturbation.
Apparently, the distinction between the pure RD and the very fast eKD$\rightarrow$RD transition is hard to be made. By decreasing the $\eta_\text{entry}$,  the differences in the amplitude and the $f$-scaling become manifest.

We note that, in our examples, the amplitude $A_0$ for the $\delta$-distribution together with the $T_\text{rh}$ value has been chosen so that either the $\Omega_\text{PBH}$ or the $\Omega_\text{IGW}$ is maximized, given the observational constraints.
Thus, these eKD scenarios can be probed in the near future by gravitational wave observatories. 
Also, a change in the $T_\text{rh}$, apart from uplifting/downlifting the $\Omega_\text{IGW}$ shifts the position of the spectrum changing the frequency $f_\text{p,IGW}$.

Regarding the energy density, the IGWs at the time $\eta_\text{c}$ in the radiation era have an energy density parameter
\begin{equation}
    \Omega^{(\text{KD})}_\text{IGW}(\eta_\text{c}) \approx \frac{\eta_\text{rh}}{\eta_\text{entry}} A_0^2\,.
\end{equation}
This result is found for a monochromatic scalar power spectrum.
Utilizing the correspondence between the $\delta$ and the Gaussian distribution with width $\epsilon$, 
\begin{equation}
\label{corresp}
A_0 \longleftrightarrow A_{\cal R}\equiv \frac{A_0}{\epsilon \sqrt{\pi}},
\end{equation}
one can find the energy density parameter of the IGWs produced during the kination era for wide ${\cal P_R}(k)$ distributions as well.

\subsection{Fitting broken power laws for IGW spectra}

In this subsection we make a few notes on the power laws that describe the IGW spectral curves in Fig.~\ref{FigIGW}, and facilitate the detectability of our results. The power laws for each case are listed in Table \ref{powerTable} together with the power-law scaling of the "mother"  curvature power spectrum.  
We separately describe the scaling in the IR part of the spectrum for $f<f_\text{p, IGW}$  and in the UV part for $f>f_\text{p, IGW}$.
For a recent independent investigation regarding the infrared scaling of the IGWs spectra see also \cite{Cai:2019cdl}, and  \cite{Kuroyanagi:2018csn} for a broken power law analysis for general stochastic GW backgrounds. 
\begin{itemize}

\item 
The Gaussian models produce IGWs that scale as a power law  $f^3$ for $f<f_\text{p,IGW}$, despite that the Gaussian ${\cal P_R}(k)$ demonstrates very steep slopes such as $k^8$ ($\epsilon=1$) or $k^{40}$ ($\epsilon=0.1$),  estimated at about the half of the curve's peak height.
In the UV frequency band, $f>f_\text{p,IGW}$, the $\epsilon=0.1$ Gaussian is almost cut off whereas the $\epsilon=1$ falls off roughly as $f^{-10}$.

\item The first $\alpha$-attractors inflationary model,
given by Eq.~(\ref{Vattrpol}),  
produces a ${\cal P_R}(k)$ that increases like $k^4$ in the beginning,  like $k^3$ before the peak and then falls like $k^{-1.2}$. The spectrum of the IGWs scales mainly $f^3$ in the IR frequencies,
 similarly to the power-law scaling of the GNMDC model. In the UV it scales  like a $f^{-2.7}$ . The IGW spectrum is found to be  broader. 

\item The second $\alpha$-attractors inflationary model, given by Eq.~(\ref{Vrun}), triggers evaporating mini-PBH formations and induced GWs near the LIGO frequency band.
The produced ${\cal P_R}(k)$ increases like $k^{4}$ and falls like $k^{-4}$. The associated IGWs scale like $f^3$ in the IR band and like $f^{-9}$ in the UV  band.

\item The Horndeski general nonminimal derivative coupling (GNMC) inflation model produces a ${\cal P_R}(k)$ that scales like  $k^{2/5}$ at small wave numbers, after like $k^2$ and then falls like $k^{-4}$. The IGW spectrum has mainly a $f^3$ scaling in the IR band and $f^{-7}$ power law scaling 
in the UV. 

\end{itemize}

Apparently, the distinction between $\alpha$-attractors and the Horndeski GNMDC inflationary models can be  made in the large-$f$ region.

\begin{itemize}
    \item  For the eKD case, the scaling of IGWs follows a power-law behavior, proportional to $f$ in the IR band. The UV part drops abruptly, consistent with our choice of delta distribution for $\mathcal{P}_{\mathcal{R}}$. In the case of a fast transition into RD, the scaling, as expected, goes like $f^2$ in the IR.
\end{itemize}

We summarize the scaling for IGWs  in Table~\ref{powerTable}.

\begin{table}[h]
    \centering
    \footnotesize{
    \begin{tabular}{|c|| c |c |c || c | c|}
    \hline
      \textbf{era} & $\boldsymbol{{\cal P_R}(k)}$  &
      $\boldsymbol{k_\text{IR}}$ &
      $\boldsymbol{k_\text{UV}}$ &
      $\boldsymbol{f_\text{IR}}$ & 
      $\boldsymbol{f_\text{UV}}$\\
    \hline \hline
          &   Gauss.  $\#1$    &   &       & $f^3$ & cut-off \\
        & Gauss.  $\#2$    &  &       & $f^3$ & $f^{-10}$ \\
 RD    &  $\alpha$-attr. $\#1$ &    $k^4, k^3$ & $k^{-1.2}$    & $f^3$ & $f^{-2.7}$ \\
       &  $\alpha$-attr. $\#2$   &  $k^4$ & $k^{-4}$  & $f^3$ & $f^{-9}$ \\
     &    GNMDC    & $k^{2/5}$, $k^2$ & $k^{-4}$          & $f^3$ & $f^{-7}$ \\
      &   $\delta$          &  - & -       & $f^2$ & cut-off \\
         \hline 
       fast eKD &  $\delta$ &  - & -   & $f^2$ & cut-off \\
     slow eKD & $\delta$    &  - & -   & $f$   & cut-off \\
     \hline 
    \end{tabular}}
    \caption{We list the approximate power-law scaling that describes the IGW spectral shapes for the ${\cal P_R}(k)$ models we studied, from IR to UV bands. $k_\text{IR}$ $(k_\text{UV})$ represents the wave numbers $k<k_\text{p}$ ($k>k_\text{p}$), and $f_\text{IR}$ $(f_\text{UV})$  the frequencies $f<f_\text{p,IGW}$ ($f>f_\text{p,IGW}$). 
    The "cutoff" means that the IGW spectrum falls off very abruptly in the UV.
    The Gauss. $\#$1 and Gauss. $\#$2 models stand for the narrow and medium width Gaussians $\epsilon=0.1$, $\epsilon=1$, respectively. The $\alpha$-attr. $\#1$ and $\alpha$-attr. $\#2$ stand for the $\alpha$-attractors inflation  models with potentials (\ref{Vattrpol}) and (\ref{Vrun}), respectively.
    }
    \label{powerTable}
\end{table}

In the IR band, as a general rule, GWs induced during the radiation era have a spectrum that follows a $f^3$ power law under for a broad $\cal P_R$ and a $f^2$ power-law for a monochromatic $\cal P_R$.\footnote{We comment that our IGW spectra are found to be broad enough and are not expected to experience a deformation, mentioned recently in \cite{Domcke:2020xmn},
via the Sachs-Wolfe and integrated Sachs-Wolfe effect as the GWs  travel toward the detectors.} Our result agrees with the conclusions of \cite{Cai:2019cdl} as well as with the GW spectrum described in \cite{Pi:2020otn}.\footnote{Note that our $\epsilon$ is  $\sqrt{2}\Delta$ in \cite{Pi:2020otn}.}
In the slow eKD$\rightarrow$RD scenario, where most of the GWs are induced during the kination era, the spectrum increases as $f$ in IR frequency band. 

In the UV frequency band, for almost all cases,
the spectrum falls off very abruptly and it is cutoff in the monochromatic cases. The exception is the first $\alpha-$attractors inflation model (\ref{Vattrpol}) that features a broader ${\cal P_R}(k)$ that falls off slowly, ${\cal P_R}(k)\sim k^{-1.2}$. Hence, broad scalar power spectra can be distinguished from the narrow ones via the spectrum of the induced GWs.

\section{Conclusions} 
\label{secConclusions}

In this work we have explored the features of the scalar-induced GW spectrum produced by  different types of ${\cal P_R}(k)$ peaks and for two different early universe cosmological scenarios: radiation and kination domination. In addition, we examined two explicit inflationary models that generate PBHs,  $\alpha$-attractors, and Horndeski general nonminimal derivative coupling models, and tested the predicted IGW signals against the observational constraints.

Assuming a Gaussian ${\cal P_R}(k)$ with amplitude $A_{\cal R}$ and width $\epsilon$ we find that that the IGWs produced during radiation era have a spectral energy density parameter today, at the frequency $f_\text{p}$ where the ${\cal P_R}$ maximizes,  given by
\begin{align} \label{OmegaRDt0}
    \Omega^{(\text{RD})}_\text{IGW}(t_0, f_\text{p}) \sim \, 5.2\times 10^{-9} \, \epsilon^2\, \left(\frac{g_*}{106.75} \right)^{-1/3} 
    \left(\frac{A_{\cal R}}{10^{-2}} \right)^2. 
\end{align}
For IGWs produced during kination domination, with $T_\text{rh}$ the reheating temperature, the corresponding expression for the energy density today is
\begin{align} \label{OmegaKint0}
    \Omega^{(\text{KD})}_\text{IGW}(t_0, f_\text{p}) \sim \, \pi \,\Omega_\text{IGW}^{\text{(RD)}}(t_0, f_\text{p}) \left(\frac{10^{7}\, \text{GeV}}{T_\text{rh}} \right) \left( \frac{f_\text{p}}{\text{Hz}}\right)
\end{align}
where $\Omega_\text{IGW}^{\text{(RD)}}(t_0, f_\text{p})$ is given by Eq. (\ref{OmegaRDt0}).

The RD  and an eKD  era may be distinguished by their power-law scaling in the small-$f$ band, $f<f_\text{p,IGW}$.
An eKD scenario with a slow reheating and sudden transition to RD predicts spectral shapes with large amplitudes for IGWs. Over the next years, aLIGO, reaching its design sensitivity, will put constraints on the eKD scenario in the high frequency band of the spectrum.
Hints for the reheating temperature can be found if the IGW spectrum has been modified due to the transition into the radiation era.

A calculation involving a $\delta$-distribution is simpler to implement, compared to the Gaussian or any other realistic distribution such as those of $\alpha$-attractors, since the integrals (\ref{tensorPSD}) can be computed analytically. We used $\delta$, i.e, monochromatic, distributions only for the eKD case, because a broad distribution is computationally more costly in that case.
Utilizing the correspondence, Eq. (\ref{corresp}), between $\delta$ and Gaussian distributions, $A_0 \longleftrightarrow A_{\cal R}$, the maximum $\Omega_\text{IGW}$ can be found following analytic steps and in a good approximation, either for the RD or the eKD case.

The spectral shape for the $\Omega_\text{IGW}(t_0, f)$ depends on the features of the source, the scalar spectrum ${\cal P_R}(k)$.
It maximizes at a  frequency $f_\text{p,IGW}$, in a little offset from $f_\text{p}$,  depending on the width of the ${\cal P_R}(k)$, as the results listed in the Tables \ref{tabRadf} and \ref{tabKinf} demonstrate. Although the power-law scaling of the IGWs increases roughly universally as  $f^3$  for RD, it falls off with different scaling.
It is interesting to mention that, for the radiation domination case at least, the ${\cal P_R}(k)$ shape is projected in a much more informative manner onto the $\Omega_\text{IGW}(f)$ spectrum than on the PBH mass distribution, which is predominantly monochromatic.
By observing the IGW spectral shape and the power law scaling in the UV frequency band one can infer the width and the amplitude of the scalar spectrum, the generator of the IGWs.  

Consequently, we can say that the detection of the $\Omega_\text{IGW}(f)$ spectrum is a portal to the primordial power spectrum of curvature perturbations, ${\cal P_R}(k)$. It can be used to discriminate inflationary models, and our analysis aimed at contributing to this direction.

\section*{Acknowledgments}
We would like to thank A. Kehagias,  C. Kouvaris and A. Riotto for discussions.
The work of I.D. is supported by IKY Scholarship, cofinanced by Greece and the European Union (European Social Fund), through the Operational Program "Human Resources Development, Education and Lifelong Learning" in the context of the project “Reinforcement of Postdoctoral Researchers - 2nd Cycle” (MIS-5033021), implemented by the State Scholarships Foundation.


\begin{table*} [h]
\footnotesize{
\begin{calstable}
\colwidths{{\colN}{\colW}{\col}{\col}{\col}{\colW}{\colW}}
\makeatletter
\def\cals@framers@width{1.4 pt}   
\def\cals@framecs@width{0.8pt}
\def\cals@bodyrs@width{0.8pt}
\cals@setpadding{Ag}
\cals@setcellprevdepth{Al}
\def\cals@cs@width{0.8pt}             
\def\cals@rs@width{0.8pt}
\def\cals@bgcolor{}

\thead{%
\bfseries
\brow
    \alignC\cell{PBH Mass} 
     \cell{\vfil $\boldsymbol{{\cal P_R}(k)}$ type}
    \cell{\vfil $\boldsymbol{A_{\cal R}}$ }
    \cell{\vfil $\boldsymbol{{\Omega_\text{PBH}}/{\Omega_\text{DM}}}$}
     \cell{ \vfil $\boldsymbol{h^2 \,\Omega_\text{IGW} }$ }
      \cell{\vfil $\boldsymbol{f_\text{p,IGW}}$  }
     \cell{\vfil \footnotesize{Experi-\\ ment}}
\erow
\mdseries
}
\brow
    \nc{lrt}
    \cell{\vfil \textbf{Gaussian 0.1}}
    \cell{$1.3\times 10^{-1}$}
     \nc{lrt}
     \cell{$9.3 \times 10^{-10}$ }
      \cell{$1.4 \times 10^{-1}$ Hz}
     \nc{lrt}
\erow
\brow
    \nc{lr}
    \cell{\vfil \textbf{Gaussian 1}}
   \cell{$4.1\times 10^{-2}$}
    \nc{lr}
      \cell{$3.3 \times 10^{-9}$ }
      \cell{$1.4 \times 10^{-1}$ Hz}
        \nc{lr}
\erow
\brow
    \nc{lr}
    \cell{\vfil \textbf{$\alpha$-attractors} $\#1$}
    \cell{$2.1\times 10^{-2}$}
   \nc{lr}   
       \cell{$1.1 \times 10^{-9}$ }
      \cell{$9.1\times10^{-2}$ Hz}
       \nc{lr}
\erow
\brow
    \nc{lrb}\sc{\vfil $\boldsymbol{10^{18}}$ \bf{g}}
    \cell{\vfil \textbf{Galileon}}
    \cell{$2.5 \times 10^{-2}$}
   \nc{lrb}\sc{\vfil${\cal O}(1)$}
  \cell{$1.1 \times 10^{-9}$ }
      \cell{$1.4\times10^{-1}$ Hz}
      \nc{lrb}\sc{\vfil DECIGO +\\ WD/Lensing}
\erow
\brow
    \nc{lrt}
    \cell{\vfil \textbf{Gaussian 0.1}}
     \cell{$1.5\times 10^{-1}$}
  \nc{lrt}
 \cell{$1.4 \times 10^{-9}$ }
      \cell{$1.2\times10^{-3}$ Hz}
       \nc{lrt}
\erow
\brow
   \nc{lrb}\sc{\vfil $\boldsymbol{10^{22}}$ \bf{g}}
   \cell{\vfil \textbf{Gaussian 1}}
     \cell{$4.8\times 10^{-2}$}
  \nc{lrb}\sc{\vfil${\cal O}(1)$}
      \cell{$4.4 \times 10^{-9}$ }
      \cell{$1.5\times10^{-3}$ Hz}
      \nc{lrb}\sc{\vfil LISA  +\\ Lensing}
\erow
\brow
    \nc{lrt}
    \cell{\vfil \textbf{Gaussian  0.1}}
     \cell{$2.2 \times 10^{-1}$}
 \nc{lrt}
    \cell{$8.9 \times 10^{-9}$ }
      \cell{$4.0 \times10^{-10}$ Hz}
         \nc{lrt}
\erow
\brow
   \nc{lrb}\sc{\vfil $\boldsymbol{10^{35}}$ \bf{g}}
   \cell{\vfil \textbf{Gaussian  1}}
    \cell{$6.2 \times 10^{-2}$}
  \nc{lrb}\sc{\vfil ${\cal O}(10^{-5})$}
     \cell{$2.6 \times 10^{-8}$ }
      \cell{$4.1\times10^{-10}$ Hz}
      \nc{lrb}\sc{\vfil PTA +\\ Lensing, X-rays}
\erow
\makeatletter
\end{calstable}\par 
}
\caption{\small{ 
The values for the  PBHs  and the associated IGWs produced during {\it radiation} domination are listed. 
The types of ${\cal P_R}(k)$ are listed together with 
the $A_{\cal R}$, the amplitude of the ${\cal P_R}(k)$ peak, and an estimation of the produced PBH abundance.
 $h^2 \Omega_{\textrm{IGW}}=h^2 \int d\ln f\ \Omega_{\textrm{IGW}}(t_0,f)$  is the present total energy density value. The $f_\text{p,IGW}$ is the frequency that the IGW energy density peaks. In the last column we quote the experiments that probe each scenario.
}}\label{tabRad}
\end{table*}

\begin{table*} [h]
\footnotesize{
\begin{calstable}
\colwidths{{\colN}{\col}{\colN}{\col}{\colN}{\colN}{\col}{\colW}}

\makeatletter
\def\cals@framers@width{1.4 pt}   
\def\cals@framecs@width{0.8pt}
\def\cals@bodyrs@width{0.8pt}
\cals@setpadding{Ag}
\cals@setcellprevdepth{Al}
\def\cals@cs@width{0.8pt}             
\def\cals@rs@width{0.8pt}
\def\cals@bgcolor{}

\thead{%
\bfseries
\brow
    \alignC\cell{PBH Mass} 
     \cell{\vfil $\boldsymbol{{\cal P_R}(k)}$ type}
    \cell{\vfil $\boldsymbol{A_0}$ }
    \cell{\vfil $\boldsymbol{{\Omega_\text{PBH}}/{\Omega_\text{DM}}}$}
      \cell{\vfil $\boldsymbol{T_\text{rh}}$ }
     \cell{ \vfil $\boldsymbol{h^2 \,\Omega_\text{IGW} }$ }
      \cell{\vfil $\boldsymbol{f_\text{p,IGW}}$  }
      \cell{\vfil \footnotesize{Experi-\\ ment}}
\erow
\mdseries
}
\brow
    \nc{lrt}
 \nc{lrt}
    \cell{$1.6\times 10^{-2}$}
\cell{${\cal O}(1)$}
  \cell{$10^{6}$ GeV}  
    \cell{$6.2\times 10^{-8}$}     
     \cell{$1.1 \times 10^{-1}$ Hz}
     \nc{lrt}
\erow
\brow
      \nc{lrb}\sc{\vfil $\boldsymbol{10^{18}}$ \bf{g}}
    \nc{lrt}
    \cell{$1.2\times 10^{-2}$}
  \cell{\vfil ${\cal O}(10^{-5})$}    
      \cell{$5 \times 10^4$ GeV}   
 \cell{$1.3\times 10^{-6}$}     
     \cell{\vfil $5.4\times 10^{-1}$ Hz}
         \nc{lrb}\sc{\footnotesize{ DECIGO \\+\\ WD/Lensing}}
\erow
\brow
    \nc{lrt}
    \nc{lrb}\sc{\vfil $\delta$-distribution}
    \cell{$1.7 \times 10^{-2}$}
\nc{lrt}
  \cell{$10^4$ GeV}
     \cell{$1.1 \times 10^{-6}$}     
     \cell{$1.0 \times 10^{-3}$ Hz}
     \nc{lrt}
\erow
\brow
   \nc{lrb}\sc{\vfil $\boldsymbol{10^{22}}$ \bf{g}}
   \nc{lr}
    \cell{$1.5 \times 10^{-2}$}
 \nc{lrb}\sc{\vfil ${\cal O}(10^{-1})$}    
    \cell{$5\times 10^2$ GeV}    
     \cell{$2.2 \times 10^{-6}$}     
     \cell{$4.2 \times 10^{-3}$ Hz}
      \nc{lrb}\sc{\footnotesize{ LISA  \\+\\ Lensing}}
\erow
\brow
   \cell{\vfil  $\boldsymbol{10^{35}}$ \bf{g}}
   \cell{\vfil }
    \cell{\vfil  $2.3 \times 10^{-2}$}
\cell{\vfil  ${\cal O}(10^{-5})$}
 \cell{\vfil  $0.01$ GeV}   
      \cell{\vfil $3.1 \times 10^{-8}$}     
     \cell{\vfil $5.5 \times 10^{-10}$ Hz}
      \cell{\vfil \footnotesize{PTA +\\ Lensing/X-rays}}
\erow
\makeatletter
\end{calstable}\par 
}
\caption{
\small{
As in Table \ref{tabRad}, the values for the  PBHs  and the associated IGWs produced during {\it kination} domination are listed.
 The $A_0$ is the coefficient of the $\delta$  distribution and $T_\text{rh}$  the  reheating temperature for each scenario.
}}\label{tabKin}
\end{table*}


\begin{center}
\begin{table*} [h]

\footnotesize{
\begin{calstable}

\colwidths{{\colN}{\colW}{\colN}{\colN}{\colN}{\col}{\colW}{\colN}}

\makeatletter
\def\cals@framers@width{1.4 pt}   
\def\cals@framecs@width{0.8pt}
\def\cals@bodyrs@width{0.8pt}
\cals@setpadding{Ag}
\cals@setcellprevdepth{Al}
\def\cals@cs@width{0.8pt}             
\def\cals@rs@width{0.8pt}
\def\cals@bgcolor{}

\thead{%
\bfseries
\brow
    \alignC\cell{\vfil $\boldsymbol{f_\text{hor}}$} 
     \cell{\vfil $\boldsymbol{{\cal P_R}(k)}$ type}
    \cell{\vfil $\boldsymbol{A_{\cal R}}$ }
    \cell{\vfil $\boldsymbol{M_\text{PBH}}$ }
    \cell{\vfil $\boldsymbol{\beta_\text{PBH}}$}
     \cell{\vfil $\boldsymbol{h^2 \,\Omega_\text{IGW} }$ }
      \cell{\vfil $\boldsymbol{f_\text{p,IGW}}$  }
       \cell{\footnotesize{Experi-\\ ment}}
\erow
\mdseries
}
\brow
    \nc{lrt}
    \cell{\vfil \textbf{Gaussian 0.1}}
    \cell{$7.8\times 10^{-2}$}
     \nc{lrt}
    \nc{lrt}
      \cell{$3.4 \times 10^{-10}$ }
      \cell{$1.1\times 10^{2} $ Hz}
     \nc{lrt}
\erow
\brow
    \nc{lr}
    \cell{\vfil \textbf{Gaussian 1}}
   \cell{$2.5\times 10^{-2}$}
    \nc{lr}
 \nc{lr}
      \cell{$1.2 \times 10^{-9}$ }
          \cell{$1.0\times 10^{2} $ Hz}        \nc{lr}
\erow
\brow
      \nc{lrb}\sc{\vfil $\boldsymbol{10^{2}}$ \bf{Hz}}
    \cell{\vfil \textbf{$\alpha$-attractors} $\#2$}
    \cell{$2.0\times 10^{-2}$}    
    \nc{lrb}\sc{\vfil $7 \times 10^{12}$ g}
   \nc{lrb}\sc{\vfil ${\cal O}(10^{-24})$}
     \cell{$1.9 \times 10^{-10}$ }
     \cell{$3.7\times 10^{1} $ Hz}
         \nc{lrb}\sc{\vfil LIGO}
\erow
\brow
    \nc{lrt}
    \cell{\vfil \textbf{Gaussian 0.1}}
    \cell{$9.0\times 10^{-2}$}
 \nc{lrt}
    \nc{lrt}
 \cell{$4.5  \times 10^{-10}$ }
       \cell{$1.1\times 10^{-1} $ Hz}
     \nc{lrt}
\erow
\brow
    \nc{lr}
    \cell{\vfil \textbf{Gaussian 1}}
   \cell{$2.7\times 10^{-2}$}
   \nc{lr}
   \nc{lr}   
 \cell{$ 1.4 \times 10^{-9}$ }
      \cell{$1.0\times 10^{-1} $ Hz}
        \nc{lr}
\erow
\brow
      \nc{lrb}\sc{\vfil $\boldsymbol{10^{-1}}$ \bf{Hz}}
    \cell{\vfil \textbf{$\alpha$-attractors} $\#2$}
    \cell{$1.4\times 10^{-2}$}
      \nc{lrb}\sc{\vfil $7 \times 10^{18}$ g}
   \nc{lrb}\sc{\vfil ${\cal O}(10^{-22})$}
     \cell{$ 5.2 \times 10^{-10}$ }
       \cell{$3.4\times 10^{-2} $ Hz}
         \nc{lrb}\sc{\vfil DECIGO}
\erow
\brow
    \nc{lrt}
    \cell{\vfil \textbf{Gaussian 0.1}}
     \cell{$8.7\times 10^{-2}$}
     \nc{lrt}
\nc{lrt}
     \cell{$4.2 \times 10^{-10}$ }
      \cell{$1.1\times 10^{-3} $ Hz}
       \nc{lrt}
\erow
\brow
   \nc{lrb}\sc{\vfil $\boldsymbol{10^{-3}}$ \bf{Hz}}
   \cell{\vfil \textbf{Gaussian 1}}
     \cell{$3.0\times 10^{-2}$}
        \nc{lrb}\sc{\vfil  $7\times 10^{22}$ g}
  \nc{lrb}\sc{\vfil ${\cal O}(10^{-20})$}
      \cell{$1.7 \times 10^{-9}$ }
       \cell{$1.0\times 10^{-3} $ Hz}
      \nc{lrb}\sc{\vfil LISA}
\erow
\brow
    \nc{lrt}
    \cell{\vfil \textbf{Gaussian 0.1}}
     \cell{$1.2 \times 10^{-1}$}
   \nc{lrt}
   \nc{lrt}
     \cell{$1.7 \times 10^{-9}$ } 
       \cell{$1.1\times 10^{-9} $ Hz}
         \nc{lrt}
\erow
\brow
   \nc{lrb}\sc{\vfil $\boldsymbol{10^{-9}}$ \bf{Hz}}
   \cell{\vfil \textbf{Gaussian 1}}
    \cell{$3.8 \times 10^{-2}$}
       \nc{lrb}\sc{\vfil$7 \times 10^{34}$ g}
\nc{lrb}\sc{\vfil${\cal O}(10^{-16})$}
      \cell{$5.9  \times 10^{-9}$ } 
        \cell{$1.0\times 10^{-9} $ Hz}
      \nc{lrb}\sc{\vfil PTA }
\erow
\makeatletter
\end{calstable}\par 
}
\caption{ \small{ The values for IGWs produced during {\it radiation} domination with zero or negligible  PBH abundance,     $\Omega_\text{PBH}/\Omega_\text{DM}\sim 10^{-10}$. The parameters are as in Table \ref{tabRad}.  The $\beta(M)$ values are listed, Eq.~(\ref{brad}), since PBHs with mass $M\lesssim 10^{15}$ g have already evaporated.
}}
\label{tabRadf}
\end{table*}
 \end{center}

\begin{table*} 
\footnotesize{

\begin{calstable}

\colwidths{{\colN}{\col}{\colN}{\colN}{\colN}{\colN}{\colN}{\col}{\colN}}

\makeatletter
\def\cals@framers@width{1.4 pt}  
\def\cals@framecs@width{0.8pt}
\def\cals@bodyrs@width{0.8pt}
\cals@setpadding{Ag}
\cals@setcellprevdepth{Al}
\def\cals@cs@width{0.8pt}           
\def\cals@rs@width{0.8pt}
\def\cals@bgcolor{}

\thead{%
\bfseries
\brow
    \alignC\cell{\vfil $\boldsymbol{f_\text{hor}}$} 
     \cell{\vfil $\boldsymbol{{\cal P_R}(k)}$ type}
    \cell{\vfil $\boldsymbol{A_0}$ }
    \cell{\vfil $\boldsymbol{M_\text{PBH}}$ }
    \cell{\vfil $\boldsymbol{\beta_\text{PBH}}$}
  \cell{\vfil $\boldsymbol{T_\text{rh}}$ }   
     \cell{\vfil $\boldsymbol{h^2 \,\Omega_\text{IGW} }$ }
      \cell{\vfil $\boldsymbol{f_\text{p,IGW}}$  }
       \cell{\footnotesize{Experi-\\ ment}}
\erow
\mdseries
}
\brow
    \nc{lrt}
\nc{lrt}    
    \cell{$9\times 10^{-3}$}
    \cell{$10^{13}$ g}
\cell{${\cal O}(10^{-25})$}
\cell{$10^9$ GeV}    
      \cell{$5.3 \times 10^{-9}$ }
       \cell{$7.1\times 10^{1} $ Hz}
     \nc{lrt}
\erow
\brow
      \nc{lrb}\sc{\vfil $\boldsymbol{10^{2}}$ \bf{Hz}}
  \nc{lrt} 
    \cell{$4\times 10^{-4}$}
      \cell{$5\times 10^{14}$ g}
    \cell{${\cal O}(10^{-550})$}
       \cell{$10^6$ GeV}
     \cell{$1.7 \times 10^{-8}$ }
       \cell{$6.2\times 10^{1} $ Hz}
         \nc{lrb}\sc{\vfil \footnotesize{LIGO}}
\erow
\brow
      \cell{\vfil $\boldsymbol{10^{-1}}$ \bf{Hz}}
   \nc{lrb}\sc{\vfil $\delta$-distribution}
    \cell{$8.8 \times10^{-3}$}
      \cell{$5\times 10^{20}$ g}
    \cell{${\cal O}(10^{-26})$}
 \cell{$10^3$ GeV}   
     \cell{$2.3 \times 10^{-6}$ }
       \cell{$3.1\times 10^{-2} $ Hz}
         \cell{\vfil \footnotesize{DECIGO}}
\erow
\brow
   \cell{\vfil $\boldsymbol{10^{-3}}$ \bf{Hz}}
 \nc{lr}
     \cell{$10^{-2}$}
       \cell{$6\times 10^{24}$ g}
 \cell{${\cal O}(10^{-24})$}
 \cell{10 GeV}    
     \cell{$2.2 \times 10^{-6}$ }
       \cell{$3.1\times 10^{-4} $ Hz}
      \cell{\vfil \footnotesize{LISA}}
\erow
\brow
   \cell{\vfil $\boldsymbol{10^{-9}}$ \bf{Hz}}
    \cell{}
    \cell{$1.5 \times 10^{-2}$}
     \cell{$ 10^{35}$ g}
\cell{${\cal O}(10^{-16})$}
 \cell{\vfil 0.02 GeV}   
     \cell{$7.7 \times 10^{-9}$}
       \cell{$1.6\times 10^{-9} $ Hz}
      \cell{\vfil \footnotesize{PTA}}
\erow
\makeatletter
\end{calstable} \par 
}
\caption{
\small{As in Table \ref{tabRadf}, the values for IGWs produced during {\it kination} domination for different reheating temperatures and with zero or negligible PBH abundance, $\Omega_\text{PBH}/\Omega_\text{DM}\sim 10^{-10}$, are listed.}}\label{tabKinf}
\end{table*}


\appendix
\section{GWs during kination era and constraints on the ${\cal P_R}(k)$ and the reheating temperature.} \label{TmaxKin}

The gravitational wave energy density gets enhanced during the kination regime \cite{Giovannini:1999bh, Riazuelo:2000fc,Yahiro:2001uh,Boyle:2007zx}.
The energy density of the GWs does not alter BBN predictions if
\begin{align} \label{I1BBN}
I \equiv h^2 \int^{k_\text{max}}_{k_\text{BBN}} \Omega_\text{GW}(k, \eta_0)d \ln k \leq  2\times 10^{-6}\,.
\end{align} 
Equivalently, the above constraint can be written in terms of the $\Omega_\text{GW}(k, \eta_\text{c})$.
The duration of the kination era is constrained due to energy density of the runaway field $\varphi$ and the GWs, see \cite{Dimopoulos:2017zvq, Akrami:2017cir, Artymowski:2017pua} for a recent discussion in this context. Here, we  derive a bound on the duration of the kination era coming from the amplitude of the IGWs, thus the ${\cal P_R}(k)$ at small scales, see also \cite{Kohri:2018awv, Domenech:2019quo}.
Let us assume that the energy of GWs is stored mainly in a narrow wave band $(k_1, k_2)$ with central wave number $k_\text{p}$ that enters the horizon at $\eta_\text{entry}$.
During kination regime the GW energy density parameter scales as $\Omega_\text{GW}\propto a^2$. At the time of reheating it is
\begin{align} \nonumber
\Omega_\text{GW}(\eta_\text{rh})
 \simeq
\frac{1}{2}\frac{\rho_\text{GW}(\eta_\text{entry})}{ \rho_\text{tot}(\eta_\text{entry})}\left(\frac{a(\eta_\text{rh})}{a(\eta_\text{entry})}\right)^2
\end{align}
where we assumed equipartition of energy densities at $T_\text{rh}$; in the case of sudden transition, the 1/2 factor should be dropped. Therefore,
$\Omega_\text{GW}(\eta_\text{c}) \simeq \Omega_\text{GW}(\eta_\text{rh})
\simeq
\frac12 \Omega_\text{GW}(\eta_\text{entry})\, ({k_\text{p}}/{k_\text{rh}})$.
Let us make the top-hat approximation for the GW spectrum, $\Omega_\text{GW}=A_\text{GW}$ in the interval $(k_1, k_2)$. Then, the integral (\ref{I1BBN}) can be estimated. Plugging in numbers we find approximately the constraint
\begin{align}
\frac{k_\text{p}}{k_\text{rh}} 
\,A_\text{GW} \,
\ln\left(\frac{k_2}{k_1}\right)
\lesssim 0.4\,.
\end{align} 
For a ballpark analytic estimation, let us assume the GW spectrum width $k_2/k_1=10$ and consider the scaling  
$\Omega_\text{GW}(\eta_\text{c})\sim A_{\cal R}^2$ during radiation domination era.

From the relation $k_\text{rh}=2\times 10^7 (T_\text{rh}/\text{GeV}) \, \text{Mpc}^{-1}$
and the kination era relation $k_\text{p}=5.4 \times 10^{28} g_*^{-1/6} (T_\text{rh}/\text{GeV})^{-1/3} (M/\text{g})^{-2/3}\gamma^{2/3}\text{Mpc}^{-1}$, see Eq.~(\ref{kMgen}),
a rough conservative lower bound on the reheating temperature for the kination regime is found,
\begin{align}
   {T_\text{rh}}\gtrsim  10^{7}\,{\text{GeV}}\, 
    A_{\cal R}^{3/2}\left(\frac{M_\text{hor}}{10^{20}\text{g}} \right)^{-1/2},
\end{align}
where $A_{\cal R}\equiv {\cal P_R}(k_\text{p})$. If this  bound is violated, the BBN predictions are endangered.

\section{Explicit PBH generating inflation models} \label{AInf}

In inflationary cosmology the primordial perturbations are produced from quantum fluctuations with a vast range of wave lenghts.
Here, we briefly present two sort of inflationary models that generate an enhanced amplitude for the ${\cal P_R}(k)$ at small scales: the $\alpha$-attractors \cite{Dalianis:2018frf, Dalianis:2019asr} and the GNMDC inflation models \cite{ Dalianis:2019vit, Fu:2019ttf}.

\subsection{$\alpha$-attractors inflation}
\begin{figure}[h]
    \centering
    \includegraphics[width=8.5cm,height=5cm]{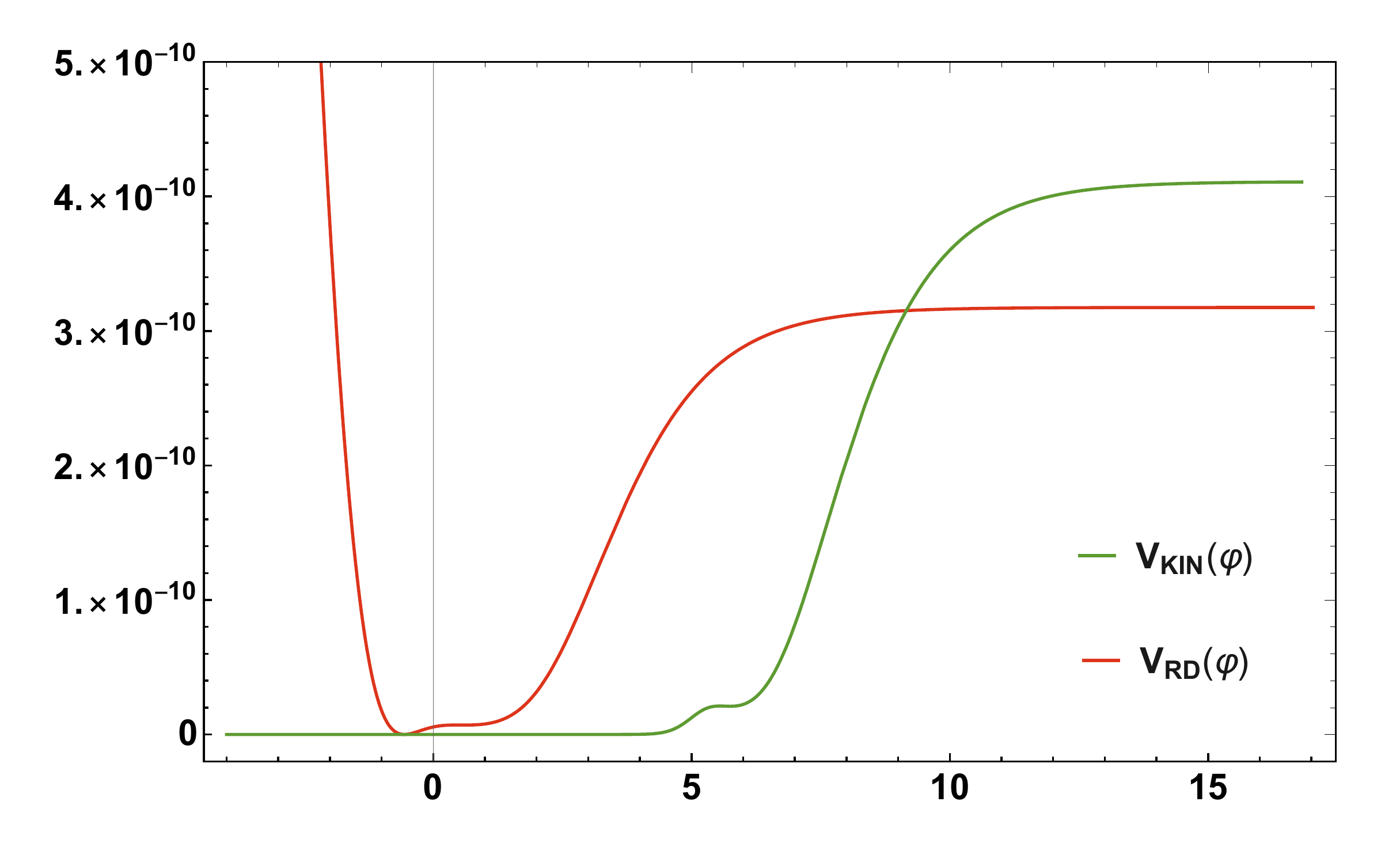}
    \caption{The figure depicts the $a$-attractors inflationary potentials that generate CMB anisotropies, enhanced small-scale perturbations that trigger PBH formation, and induced GWs. A reheating phase
    follows inflation for the potential in red and a kination phase for the potential in green. The amplitude of the potential in green has been adjusted to fit in the same plot. }
     \label{Figaattrac}
\end{figure}

In the $\alpha$-attractors inflation scenario \cite{Kallosh:2013yoa}, the effective Lagrangian for the inflaton field $\varphi$  turns out to be 
\begin{equation} \label{la11}
\frac{\cal L}{\sqrt{-g}}=  \frac{1}{2}R-\frac{1}{2}
\Big(\partial_\mu \varphi\Big)^2-f^2\Big(\tanh\frac{\varphi}{\sqrt{6\alpha}}\Big)\,,
\end{equation}
where Re$\Phi=\sqrt{3}\tanh({\varphi}/{\sqrt{6\alpha}})$ is a chiral superfield, and $R$ is the Ricci scalar. We also took $M_\text{Pl}=1$.
The canonically normalized inflaton potential that determines the   inflationary trajectory is
$V(\varphi)=f^2(\tanh\varphi/\sqrt{6\alpha})$. 
Polynomial, trigonometric and exponential forms for the function $f(\text{Re}\Phi)$ can feature   an inflection point plateau sufficient to generate a significant dark matter abundance in accordance with the observational constraints  \cite{Dalianis:2018frf}.
In this work we presented tensor power spectra induced by scalar power spectra ${\cal P_R}(k)$ predicted by the inflationary potentials
\begin{equation}\label{Vattrpol}
V(\varphi)=f^2_0 \left[\sum_{n=0}^3 c_n \left(\tanh({\varphi}/{\sqrt{6\alpha}}\right)^n \right]^2
\end{equation}
and
\begin{align} \label{Vrun}  \nonumber
 V(\varphi)=  f^2_0  \, \big[ c_0 + &  c_1 e^{\lambda_1 \tanh \varphi/\sqrt{6}} \, + 
 \\
&  c_2   e^{\lambda_2 \left(\tanh(\varphi/\sqrt{6})-\tanh(\varphi_\text{P} /\sqrt{6})\right)} \big]^2\,.
\end{align}
The form of the potentials (\ref{Vattrpol}) and (\ref{Vrun}) is depicted in Fig. \ref{Figaattrac}. The inflection point plateau is the field region where the curvature perturbations get enhanced. 
The potential (\ref{Vattrpol}) implements an RD cosmological postinflationary phase with high reheating temperature together with large amplitudes for ${\cal P_R}(k)$ that induce GWs. 
The potential (\ref{Vrun}) implements an inflationary phase followed by a kination phase. The kination phase can end gradually, as in the original proposal \cite{Dalianis:2019asr}, or via a sudden transition due to e.g. an extra field direction toward a global minimum.
The power spectra generated are depicted in Fig.~\ref{Figps}. 
Details about the range of the parameter values for the inflationary models (\ref{Vattrpol}) and (\ref{Vrun}) as well details about the reheating temperature and the ${\cal P_R}(k)$ amplitude and the PBH abundances can be found in the works \cite{Dalianis:2018frf} and \cite{Dalianis:2019asr}.

\begin{figure}[h]
    \centering
    \includegraphics[width=8.5cm,height=5cm]{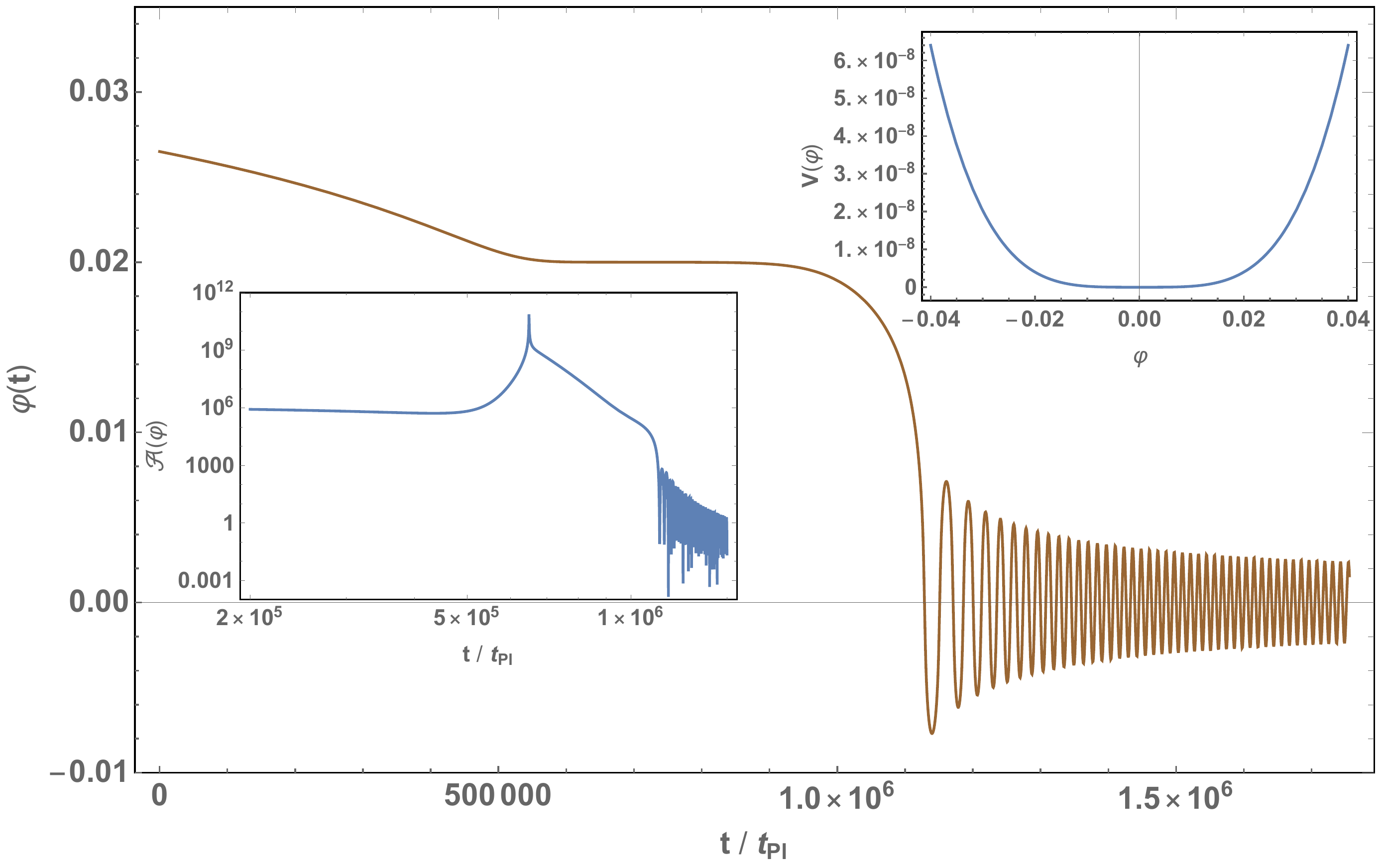}
    \caption{The main plot depicts the evolution of the inflaton field with GNMDC. 
    The inner plots depict a Higgs-like potential that together with the field-dependent GNMDC, $A(\varphi)=1+3H^2f(\varphi)$ generates a peak in the ${\cal P_R}(k)$, a PBH abundance,  and induced GWs.
    }
    \label{FigHornInf}
\end{figure}

\subsection{Horndeski GNMDC inflation}
Horndeski theories are scalar-tensor theories of gravity \cite{Horndeski:1974wa, Kobayashi:2019hrl}.
The Horndeski inflation model that we examine 
is the one with the 
nonminimal derivative coupling of the scalar field to the Einstein tensor,
with a general $\varphi$-dependent form for the function $G_5(\varphi, X)=\hat{f}(\varphi)$ that is part of the the generalized Galileon action \cite{Deffayet:2011gz}.  This term was named  general non-minimal derivative coupling (GNMDC) in \cite{Dalianis:2019vit}. The inflaton action is given by Eq.~(\ref{Lhorn})
and the potential that we use is
\begin{align} \label{Higgs}
    V(\varphi)=\frac{\lambda}{4} \varphi^4
\end{align}
that can be identified with the Higgs \cite{Dalianis:2019vit}. The form of the GNMDC is
\begin{align} \label{gnmdc}
\hat{f}(\varphi) =\frac{\alpha  \varphi^{\alpha-1}}{M^{\alpha+1}}\left( 1 + f_{II}(\varphi)\right)~,
\end{align}
with $f_{II}=d(\left((\varphi-\varphi_0)/s \right)^2+1)^{-1/2}$ that features a sharp peak at $\varphi_0$ \cite{Fu:2019ttf}.
The parameters $s$ and $d$ determine, respectively, the width and the amplitude of the $f_{II}$ term.
 The GNMDC inflation features a noncanonical kinetic term that acts like friction, decelerating the inflaton field about the $\varphi_0$ value, while the field rolls down the $\lambda \varphi^4$ potential. The form of the GNMDC enhances significantly the ${\cal P_R}(k)$ and triggers PBH formation \cite{Dalianis:2019vit}. The evolution of the inflaton scalar field  with a Higgs-like potential  is depicted in Fig. \ref{FigHornInf} when a  field-dependent GNMDC, called ${\cal A}(\varphi)$, acts.

\newpage

\bibliography{paper}

\end{document}